\documentclass[journal,twoside,comsoc]{IEEEtran}

\usepackage{booktabs}
\usepackage{ragged2e}
\usepackage{graphicx}
\usepackage{multirow}
\usepackage{float}
\usepackage[moderate]{savetrees}
\usepackage{pdflscape}
\usepackage{adjustbox}
\usepackage{array}
\usepackage{soul, color}
\usepackage{xcolor}
\usepackage{array}
\usepackage{balance}
\usepackage[english]{babel}
\usepackage{tikz}
\usetikzlibrary{arrows.meta, positioning, shapes, calc}
\usepackage{amsmath}
\usepackage{tikz}
\usepackage[numbers]{natbib}
\usepackage{tcolorbox}

\usepackage[colorlinks=true, linkcolor=black, citecolor=black, urlcolor=black]{hyperref}

\usepackage{xurl}
\usepackage{orcidlink}

\def\checkmark{\tikz\fill[scale=0.4](0,.35) -- (.25,0) -- (1,.7) -- (.25,.15) -- cycle;}

\newcommand{\journalColorText}{\textcolor{black}}

\newcounter{lessonscounter}

\newcommand{\insights}[1]{%
    \refstepcounter{lessonscounter}%
    \begin{tcolorbox}[
        colframe=black!50,
        colback=gray!10,
        coltext=black,
        sharp corners=southwest,
        boxrule=0.8pt,
        before skip=8pt,
        after skip=8pt,
        top=4pt,
        bottom=4pt,
        left=6pt,
        right=6pt,
        boxsep=1pt
    ]
        \emph{Insight \thelessonscounter: #1}
    \end{tcolorbox}
}

\newcommand{\raisedorcid}[1]{\raisebox{0.5ex}{\orcidlink{#1}}}

\begin{document}

\title{Stablecoins: Fundamentals, Emerging Issues, and Open Challenges}

\author{Ahmed Mahrous\raisedorcid{0000-0003-4694-5336},
Maurantonio~Caprolu\raisedorcid{0000-0001-8237-0539
},~\IEEEmembership{Member,~IEEE,}
and~Roberto~Di Pietro\raisedorcid{0000-0003-1909-0336},~\IEEEmembership{Fellow,~IEEE}

\thanks{The authors are with the Computer, Electrical and Mathematical Sciences and Engineering Division, King Abdullah University of Science and Technology, Thuwal,
Saudi Arabia.}%
}


\maketitle

\begin{abstract}
Stablecoins, with a capitalization exceeding 200 billion USD as of January 2025, have shown significant growth,
with annual transaction volumes exceeding 10 trillion dollars in 2023 and nearly doubling that figure in 2024.
This exceptional success has attracted the attention of traditional financial institutions, with an increasing number of governments exploring the potential of Central Bank Digital Currencies (CBDCs). Although academia has recognized the importance of stablecoins, research in this area remains fragmented, incomplete, and sometimes contradictory. 

In this paper, we aim to address the cited gap with a structured literature analysis, correlating recent contributions to present a picture of the complex economic, technical, and \journalColorText{regulatory} aspects of stablecoins. To achieve this, we formulate the main research questions and categorize scientific contributions accordingly, identifying \journalColorText{main results}, data sources, methodologies, and open research questions.

\journalColorText{The research questions we address in this survey paper cover several topics, such as the stability of various stablecoins, novel designs and implementations, and relevant regulatory challenges. The studies employ a wide range of methodologies and data sources, which we critically analyze and synthesize. Our analysis also reveals significant research gaps, including limited studies on security and privacy, underexplored stablecoins, unexamined failure cases, unstudied governance mechanisms, 
and the treatment of stablecoins under financial accounting standards, among other areas.}

\end{abstract}


\section{Introduction}
\label{sec:introduction}
A stablecoin is a class of digital assets designed to maintain a stable value relative to a reference asset or basket of assets. \journalColorText{These can be a fiat currency (such as USD), a commodity (such as gold), or other financial assets.} Stablecoins achieve price stability through various mechanisms, including full or partial collateralization, algorithmic control of supply, or a combination thereof. 
Stablecoins play an essential role in the digital finance ecosystem, addressing one of the primary limitations of cryptocurrencies---price volatility---while leveraging the benefits of blockchain technology. Their significance is highlighted by their substantial market capitalization and widespread use. The combined market capitalization of stablecoins surpasses 
\$200 billion as of the time of this 
writing~\cite{defillama_stablecoins}, a more than thirty-fold increase since the start of 2020 \cite{coingecko2024}, as shown in Fig.~\ref{fig:marketcap}. 
As for 2024, volume amounted to $8.5$ trillion across $1.1$ billion transactions in the second quarter of 2024 (ending June 2024). 
\journalColorText{Corporations, including PayPal \cite{paypal_stablecoin}, JPMorgan Chase \cite{jpmorgan_stablecoin}, and Facebook \cite{facebook_diem} have proposed or developed their own stablecoin.} Governments also showed interest. According to the Atlantic Council’s CBDC Tracker, 44 countries are piloting central bank digital currencies (CBDCs), including major economies such as China and the Eurozone \cite{atlanticcouncil_cbdc_tracker}. Most CBDC initiatives resemble stablecoins in design.

Despite the growing importance of stablecoins, reflected in academia through the increasing volume of the literature on the topic, research in this area is still in its early stages, with fragmented results and sometimes questionable methodologies.
Stablecoin research explores a diverse array of topics, reflecting the interplay of economic, technical, and \journalColorText{regulatory} aspects.

In terms of economic research, significant efforts have been dedicated to understanding the price dynamics of stablecoins, their interconnectedness with other assets and external forces, \journalColorText{their behavior during market stress, and their level of stability. Researchers also investigated the factors driving stablecoin adoption, their current and potential use cases, and the broader implications of their adoption.} Meanwhile, technical studies have explored novel design proposals, stabilization and \journalColorText{governance mechanisms}, causes and impacts of failures, and security and privacy aspects. \journalColorText{Regulatory responses and challenges have also been addressed}. The interdisciplinary nature of stablecoin research, integrating concepts from computer science, finance, and related fields, is reflected in the wide range of methodologies and data sources it employs. Methodologies include, among others, econometric models (such as regression, GARCH, and VAR models), formal theoretical modeling, network analysis, review of stablecoin white papers, and system design frameworks. Given the recency of stablecoin research, significant research gaps exist. These include gaps in critical areas such as the privacy and security of stablecoins, unexamined failure cases, the analysis of underexplored unique stablecoins, and unstudied governance mechanisms. Data sources include, among others, cryptocurrency price and trading volume data, blockchain transaction data from explorers such as Etherscan, and \journalColorText{data from DeFi platforms like Curve Finance, Convex Finance, and Votium}. 

The fragmented nature of stablecoin studies and the diversity of issues addressed make it challenging to synthesize insights. Our paper aims to fill this gap. Existing surveys often focus on particular stablecoin projects (mainly investigating their stabilization mechanisms) \journalColorText{rather than reviewing academic studies at large}. This contrasts with our approach, which seeks to synthesize current literature and guide future research. While two literature reviews \cite{ante2023systematic, li2024stablecoin} offer valuable insights, they are limited in scope and depth. For instance, \citet{ante2023systematic} reviewed only 22 empirical studies, and \citet{li2024stablecoin} focused on three predefined questions about stabilization mechanisms and payment applicability rather than a more comprehensive analysis of stablecoin literature.

\textbf{Contributions:}
Our paper contributes to the field by systematically categorizing the stablecoin literature, summarizing and analyzing key findings, methodologies, and data sources, and identifying significant open research questions.
Our main contributions can be summarized as follows.

\begin{itemize}
    \item We identify \journalColorText{26} main research questions in stablecoin-related literature and categorize existing studies accordingly;
    \item \journalColorText{We offer a comprehensive analysis of existing stablecoin research, covering and connecting the interdisciplinary dimensions of economic, technological, and regulatory aspects. By doing so, we aim to address the fragmented nature of current literature.}
    \item We summarize findings from the literature, simplifying complex analyses and highlighting areas of agreement and disagreement among researchers;
    \item After critically reviewing methodologies and conclusions, we identify significant research gaps, areas of contention, and new research questions;
    \item \journalColorText{We present original data analyses and visualizations to illustrate and support key insights of the systematic literature review.}
\end{itemize}







\textbf{Roadmap:} The rest of this paper is organized as follows. Section~\ref{sec:related_work} presents existing survey papers related to stablecoins, highlighting their limitations. Section~\ref{sec:methodology} details the methodology of our paper, including the search strategy, research categorization, and objectives. In Sections~\ref{sec:economic_analysis}, \ref{sec:technical_analysis}, and \journalColorText{\ref{sec:policy_analysis}}, we present the main findings, methodologies, data sources, and research gaps identified in papers addressing economic, technical, and policy-related aspects, respectively. Section~\ref{sec:discussion} discusses several significant findings and their implications, commonly used data sources, and overarching research gaps. Section~\ref{sec:conclusion} concludes, summarizing the main contributions of this survey paper.

\section{Related Work}
\label{sec:related_work}

\begin{table*}[htbp]{\tiny}
\renewcommand{\arraystretch}{1.3}
\caption{Comparison of Stablecoin Surveys}
\label{tab:related_work}
\centering
\begin{tabular}{|
  >{\raggedright\arraybackslash}p{2cm}|
  >{\raggedright\arraybackslash}p{2cm}|
  >{\raggedright\arraybackslash}p{2cm}|
  >{\raggedright\arraybackslash}p{2cm}|
  >{\raggedright\arraybackslash}p{2cm}|
  >{\raggedright\arraybackslash}p{2cm}|
  >{\raggedright\arraybackslash}p{2cm}|
}
\hline
\textbf{Paper} & \textbf{Systematic Literature Review} & \textbf{Categorization of Literature} & \textbf{Comprehensive Overview of Literature} & \textbf{Research Gaps Identified} & \textbf{Methodologies Identified} & \textbf{Data Sources Identified} \\
\hline
\cite{zhu2022survey} & \text{\sffamily X} & \text{\sffamily X} & \text{\sffamily X} & \checkmark & \text{\sffamily X} & \text{\sffamily X} \\
\hline
\cite{clark2019sok} & \text{\sffamily X} & \text{\sffamily X} & \text{\sffamily X} & \text{\sffamily X} & \text{\sffamily X} & \text{\sffamily X} \\
\hline
\cite{pernice2019monetary} & \text{\sffamily X} & \text{\sffamily X} & \text{\sffamily X} & \checkmark & \text{\sffamily X} & \text{\sffamily X} \\
\hline
\cite{mita2019stablecoin} & \text{\sffamily X} & \text{\sffamily X} & \text{\sffamily X} & \text{\sffamily X} & \text{\sffamily X} & \text{\sffamily X} \\
\hline
\cite{moin2020sok} & \text{\sffamily X} & \text{\sffamily X} & \text{\sffamily X} & \checkmark & \text{\sffamily X} & \text{\sffamily X} \\
\hline
\cite{dionysopoulos2024} & \text{\sffamily X} & \text{\checkmark} & \text{\sffamily X} & \checkmark & \text{\sffamily X} & \text{\sffamily X} \\
\hline
\cite{ante2023systematic} & \checkmark & \checkmark & \text{\sffamily X} & \checkmark & \checkmark & \checkmark \\
\hline
\cite{li2024stablecoin} & \checkmark & \checkmark & \text{\sffamily X} & \checkmark & \text{\sffamily X} & \text{\sffamily X} \\
\hline
\textbf{This  Paper} & \checkmark & \checkmark & \checkmark & \checkmark & \checkmark & \checkmark \\
\hline
\end{tabular}
\end{table*}


\journalColorText{Despite the increasing importance of stablecoins, reflected in a growing body of academic literature, there is a notable gap in systematic literature reviews that offer a comprehensive and in-depth overview of the field. Most existing contributions tend to focus either on classifying and describing stablecoin projects (e.g., based on their white papers) or address a limited set of research questions, without providing an analysis of the broader academic literature. Filling this gap is crucial for researchers to gain a comprehensive understanding of the stablecoin landscape, identify mature research areas and underexplored questions, and for policymakers to make informed decisions, etc.}

\journalColorText{Early work offered a valuable contribution by introducing the concept of stablecoins and categorizing them based on designs and stability mechanisms. The primary goal of these papers was mainly to present the variety of stablecoin design approaches and discuss general characteristics.}

\journalColorText{\citet{pernice2019monetary} aimed to systematically explore different stablecoin projects and connect their designs with concepts from traditional monetary policy. Their methodology involved surveying 24 stablecoin projects by analyzing their white papers, websites, and available price data. They presented a classification of stablecoins based on stabilization techniques (such as collateralization, interest rates, and open market operations) and exchange rate regimes (like hard pegs, soft pegs, and floating), highlighting key insights such as the potential vulnerability of soft pegs to speculative attacks.}

\journalColorText{\citet{mita2019stablecoin} focused on comparing stablecoin price stabilization mechanisms. Their methodology involved classifying mechanisms by four collateral types (fiat, crypto, commodity, non-collateralized) and distinguishing between mechanisms operating at the blockchain protocol level (protocol layer; e.g., mining rewards) and those built on top using smart contracts (application layer; e.g., supply rebase). They argued that non-collateralized stablecoins hold promise for decentralization, though they acknowledged their stabilization challenges.}

\journalColorText{\citet{clark2019sok} presented a Systematization of Knowledge (SoK) with the primary goal of making stablecoins accessible to a broader audience. Their methodology involved analyzing 185 articles from CoinDesk and project documentation. A key contribution was their classification of stablecoin mechanisms into ``backed'' (or collateralizated) and ``intervention-based'' (or algorithmic) approaches, alongside a discussion of the related trust dynamics (e.g., trusting a custodian versus an algorithm). They also analyzed historical market data to illustrate volatility patterns under different stabilization mechanisms and identified regulatory uncertainty as a major barrier to stablecoin adoption.}

\journalColorText{Both \citet{moin2020sok} and \citet{zhu2022survey} developed classifications based on project white papers and technical reports, categorizing stablecoins along several dimensions: peg type (fiat, crypto, algorithmic), collateral model (asset-backed, crypto-backed, algorithmic), stabilization mechanism, oracle source, governance structure, and regulatory compliance features. \citet{zhu2022survey} highlighted the inconsistent and often conflicting definitions found in stabecloin-related literature.}

\journalColorText{While these contributions offer valuable insights into proposed stablecoin designs and underlying concepts, they primarily focus on project classification or provide broad overviews rather than conducting a systematic survey of academic literature.}

\journalColorText{Moving beyond that, some papers have started to systematically review the academic literature itself, although often with a limited scope. \citet{ante2023systematic} conducted a systematic literature review focusing specifically on empirical research on stablecoins. Their primary goal was to review the current state of empirical literature across three main topic (stability/volatility, interrelations with crypto-markets, and impact of macroeconomic factors). Based on a sample of only 22 empirical studies, they analyzed the methodological approaches, data sources, and independent variables used, and mentioned future research ideas. While offering a valuable contribution, their scope was limited in terms of the number of papers analyzed and the range of topics covered, excluding theoretical, design-focused, and regulatory studies, for instance.}

\journalColorText{\citet{li2024stablecoin} provided a survey reviewing the literature related to the structure of stablecoins, stabilization mechanisms, and applicability of stablecoins as a payment system. Key insights included classifying stablecoins by collateral type (asset-backed, cryptocurrency-backed, algorithm-backed), discussing their system architecture and various stabilization mechanisms, and discussing stablecoins in the context of payment methods and in relation to the monetary impossible trinity theory. Their focus remained on these predefined topics, and their analysis did not include a detailed breakdown of specific methodologies and data sources used across stablecoin research, nor a detailed identification of research gaps. In contrast, our study provides a more comprehensive analysis, examining a broader body of stablecoin-related literature, and discussing methodologies, data sources, and research gaps. Our approach results in the coverage of topics not addressed by \citet{li2024stablecoin}, such as price modeling of stablecoins, interconnectedness with other assets, and failures of stablecoins, while also providing a more detailed coverage of those topics included in their work.}

\journalColorText{\citet{dionysopoulos2024} provided a survey of economic stablecoin research, organizing literature around asset interconnectedness, peg stability, money-market analogues, and regulatory issues. They document key trends such as the explosive market growth of stablecoins, USDT/USDC dominance, and stablecoin issuers' ranking in top 20 U.S. debt holders. They also note a general lag in academic work on stablecoins compared to the general cryptocurrency field. Our work goes significantly beyond that. We expand the scope to a more interdisciplinary analysis, integrating technical, economic, and regulatory research. While \citet{dionysopoulos2024} offered general research directions, we detail 26 distinct research questions, offering a more concrete view of past stablecoin research directions. Furthermore, we evaluate and compare the diverse methodologies and data sources—--a contribution missing in their work. Finally, compared to their general recommendation for future research, we detail concrete research gaps for every one of the 26 questions.}

\journalColorText{\citet{mahrous2025sok} presented an earlier survey of stablecoin literature which this, more comprehensive, study now extends. The primary aim of \cite{mahrous2025sok} was to provide a structured analysis of certain economic and technical aspects of stablecoin research by systematically reviewing related academic contributions. Its methodology involved identifying 13 main research questions within these domains, categorizing existing academic studies accordingly, and then synthesizing findings, critically reviewing common data sources and methodologies, and discussing related research gaps. Our current study significantly expands upon this initial work. While the conference paper analyzed 111 papers to identify 13 research questions across economic and technical topics, this paper examines all Scopus-indexed stablecoin-related literature (209 documents). This broader base enables the introduction of a more comprehensive list of 26 distinct research questions, organized into Economic Analysis, Technical Design, and the newly added category of Position/Policy Papers. Consequently, this examined work examines several research areas not previously examined, including factors impacting stablecoin adoption and its implications (e.g., economic and geopolitical impacts), design of CBDCs, stablecoin governance mechanisms, modeling and forecasting of stablecoin prices, and several regulatory issues. Moreover, the present work incorporates additional data analyses and visualizations to illustrate and support insights.}

\journalColorText{Ultimately, this study distinguishes itself from prior work by moving beyond project-centric classifications, general overviews, or more limited analyses, resulting in a more comprehensive, systematic, and interdisciplinary examination of academic stablecoin-related literature. To highlight several aspects of our contribution, Table~\ref{tab:related_work} compares this paper with previous surveys across six dimensions.}

\section{Methodology, Categorization, and Objectives}
\label{sec:methodology}
Our paper aimed to provide a systematic and comprehensive review of the academic literature related to stablecoins. To do that, we queried the Scopus Search Engine to identify papers that mention stablecoin-related keywords in the title, abstract, or keywords. This led us to a large and broad body of Scopus-indexed stablecoin-related papers, \journalColorText{that we integrated} with searches from Google Scholar \journalColorText{to include novel contributions from peer-reviewed venues---not yet indexed on Scopus---and potentially other relevant documents from trusted sources, e.g., technical reports from the European Central Bank (ECB)}. \journalColorText{This search yielded a total of 236 documents.}

The query used to search for papers on the Scopus Search Engine is the following:

\begin{quote}
\raggedright
\texttt{TITLE-ABS-KEY (stablecoin* OR "stable coin*") LOAD-DATE < 20240917 AND (LIMIT-TO (DOCTYPE, "ar") OR LIMIT-TO (DOCTYPE, "cp") OR LIMIT-TO (DOCTYPE, "re") OR LIMIT-TO (DOCTYPE, "cr")) AND (LIMIT-TO (LANGUAGE, "English"))}
\end{quote}
Some documents have been removed due to being inaccessible, i.e., behind paywalls, no longer available in their publisher digital library, etc. (10 papers), not being in English (3 papers), or not focusing significantly on Stablecoins (14 papers). The remaining \journalColorText{209} documents have been analyzed to determine their topic of analysis. \journalColorText{The research directions related to stablecoins have been systematically categorized for a comprehensive overview. A total of 26 distinct research questions have been identified and grouped into 10 general categories, which are further organized into 3 macro-categories. The macro-categories, sub-categories, and research questions are listed in Table~\ref{tab:categories}, with the number of papers in parentheses.}


\journalColorText{\textbf{Structure of the Paper:}}  
This paper provides a systematic overview of the research questions addressed in stablecoin-related academic literature. 
%
While connections between research questions are discussed where relevant, each question is analyzed in depth by correlating results, methodology, and lessons learned from all the papers that contributed significantly to addressing this particular question. 
Insights derived from this analysis are integrated throughout the discussion, with research gaps identified and addressed at the end of each sub-category. This structure allows readers to easily locate the findings, methodologies, and research gaps most relevant to their interests, minimizing search efforts. For convenience, Table~\ref{tab:categories} presents a broad overview of all discussed research questions.


\begin{figure}[htbp] 
    \centering
    \includegraphics[width=0.48\textwidth]{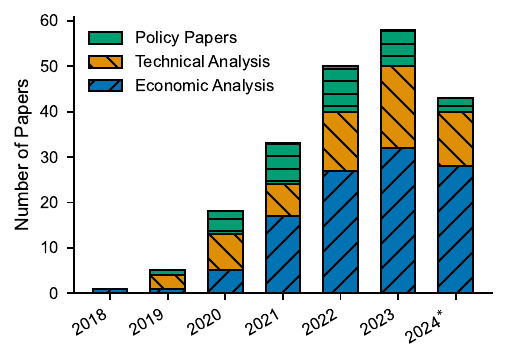}
    \caption{\journalColorText{Number of papers per year for each macro-category. The data for 2024 are updated as of the time of our search query, conducted in September 2024.}}
    \label{fig:research_trend}
\end{figure}

\journalColorText{Fig.~\ref{fig:research_trend} illustrates the number of papers per year in our sample, categorized by macro category. Note that the search query was conducted in September 2024, resulting in an incomplete count for 2024, as not all papers published in that year were included in the sample.}
\begin{table*}[htbp]
    \caption{Macro-Categories, Sub-Categories, and Research Questions. Number of related papers is in parentheses.}
    \label{tab:categories}
    \centering
    \renewcommand{\arraystretch}{1.5}
\begin{tabular}{|
  >{\raggedright\arraybackslash}p{2cm}|
  >{\raggedright\arraybackslash}p{3cm}|
  >{\raggedright\arraybackslash}p{12cm}|
}
        \hline
        \textbf{Macro-Category} & \textbf{Sub-Category} & \textbf{Research Questions} \\
        \hline
        \multirow{12}{2cm}{Economic Analysis (111)} 
         & \multirow{4}{3cm}{Stablecoin Price/Volume Correlation (61)} 
            & How does the price of stablecoins move in relationship to the price of other financial assets? (37) \\
        \cline{3-3} 
         &  & How do macroeconomic, blockchain, media, social, climate, and regulatory factors interact with stablecoins? (11)\\
        \cline{3-3} 
         &  & \journalColorText{How do stablecoin issuances/transfers impact non-stable cryptocurrencies? (8)}\\
        \cline{3-3} 
         &  & \journalColorText{How do external crises impact stablecoins? (5)}\\
        \cline{2-3} 
         & \multirow{4}{3cm}{\journalColorText{Stablecoin Usage and Adoption (27)}} 
         & \journalColorText{What are the economic, financial, and policy impacts of adopting stablecoins? (4)}\\
        \cline{3-3} 
         &  & \journalColorText{What factors influence the adoption of stablecoins? (12)}\\
        \cline{3-3}
         &  & \journalColorText{What are current use cases of stablecoins? (7)}\\
        \cline{3-3} 
         &  & \journalColorText{What are the potential use cases for stablecoins across different industries? (4)}\\
        \cline{2-3} 
         & \multirow{2}{3cm}{Stablecoin Stability Analysis (12)} 
            & How stable are stablecoins in general market conditions? (7)\\
        \cline{3-3} 
         &  & How stable are stablecoins during periods of market turbulence or crisis? (5)\\
        \cline{2-3} 
         & \multirow{3}{3cm}{Price Modeling (11)} 
            & What are the properties of stablecoins’ prices in terms of linearity, time-variation, regime-switching, normality, and autoregressive behavior, etc.? (4)\\
        \cline{3-3} 
         &  & How can actors in stablecoin ecosystems be modeled using computational simulations? (2)\\
        \cline{3-3}
         &  & \journalColorText{What are the most effective methods for modeling and forecasting stablecoin prices? (5)}\\        
        \hline
        \multirow{10}{2cm}{Technical Analysis (62)} 
         & \multirow{3}{3cm}{Design and Implementation (29)} 
            & How can we improve the design and implementation of stablecoins? (13)\\
        \cline{3-3} 
         &  & \journalColorText{How are CBDCs designed, and how should they be designed? (8)}\\
         \cline{3-3}
         & & \journalColorText{How can certain stablecoin-related services be designed and implemented? (8)}\\
        \cline{2-3} 
         & \multirow{2}{3cm}{Stablecoin Mechanisms (21)} 
            & What are the characteristics of stabilization mechanisms used in stablecoins? (14)\\
        \cline{3-3} 
         &  & \journalColorText{What are the characteristics of governance mechanisms in stablecoins, and how do they impact the platforms? (7)}\\
        \cline{2-3} 
         & \multirow{3}{3cm}{Failure Analysis (8)} 
            & What are the causes of previous stablecoin failures? (4) \\
        \cline{3-3} 
         &  & What are the impacts of previous stablecoin failures? (2)\\
        \cline{3-3} 
         &  & \journalColorText{How can we better forecast stablecoin failures? (2)}\\
        \cline{2-3} 
         & \multirow{2}{3cm}{Security and Privacy (4)} 
            & What are the privacy-concerns related to CBDCs and how are they being addressed? (1)\\
        \cline{3-3} 
         &  & What are the security threats related to CBDCs and stablecoins? (2)\\ 
        \cline{3-3}
         &  & How can market manipulation related to stablecoins be detected? (1) \\
        \hline
        \multirow{3}{2cm}{\journalColorText{Position and Policy Papers (36)}} 
         & \journalColorText{Position Papers: Overview and Implications (25)} & \journalColorText{What are the general characteristics and implications of stablecoins and CBDCs? (25)} \\ 
        \cline{2-3} 
         & \multirow{2}{3cm}{\journalColorText{Legal and Regulatory (11)}} 
            & \journalColorText{What are the regulatory challenges related to stablecoins and CBDCs, how have regulators responded, and how should these challenges be dealt with? (11)}\\ 
        \cline{3-3} 
        \hline
    \end{tabular}
\end{table*}


\section{Economic Analysis}
\label{sec:economic_analysis}

Stablecoins represent an inherently interdisciplinary topic, spanning economic, technological, and regulatory dimensions. This section specifically focuses on the economic analysis of stablecoins, synthesizing insights from 111 academic studies. Notably, the correlation of stablecoin prices and volumes with other markets and external factors emerges as the most extensively studied sub-category, highlighting stablecoins' potentially significant integration into global financial systems.

\subsection{Stablecoin Price/Volume Correlation With Other Factors}
\textit{\textbf{(1) How Does The Price Of Stablecoins Move In Relationship To The Price Of Other Financial Assets?}}

Stablecoins demonstrate potential as risk management tools due to their generally low or negative correlation with both cryptocurrencies and traditional assets, especially during periods of market stress. Their relative stability allows them to act as safe havens, though this varies by type of stablecoin. Additionally, the extent of their effectiveness as safe havens and the presence of spillover effects remain subject to ongoing debate.

This stability is especially evident during periods of high market volatility, such as the COVID-19 pandemic and geopolitical events like the Russia-Ukraine war. During these times, many financial assets experience heightened volatility, whereas stablecoins tend to maintain relative stability \cite{wang2020stablecoins, baur2021crypto, kliber2022looking}. This low or negative correlation with other assets enables stablecoins to act as a hedge against market downturns, offering investors a means to reduce portfolio risk \cite{anisiuba2021analysis, gadi2022analyzing, kolodziejczyk2023stablecoins}.

However, the effectiveness of stablecoins as safe havens or hedging instruments depends significantly on the type of stablecoin and prevailing market conditions. Gold-backed stablecoins like PAX Gold (PAXG) and Perth Mint Gold Token (PMGT) serve as strong diversifiers during normal market conditions, and as robust safe havens during bear markets for certain asset classes \cite{belguith2024can}. In contrast, USD-pegged stablecoins, such as Tether (USDT) and USD Coin (USDC), outperform gold-pegged stablecoins in reducing portfolio risk during periods of financial distress \cite{wang2020stablecoins, diaz2023stablecoins, feng2024stablecoins}. This can be attributed to the higher volatility of gold compared to the USD \cite{wang2020stablecoins} and the specific mechanisms of certain stablecoins, such as Digix Gold \cite{diaz2023stablecoins}. Additionally, USD-pegged stablecoins benefit from larger market capitalization and more reputable issuers.

\insights{There is a consensus among researchers that stablecoins typically exhibit low or negative correlation with both cryptocurrencies and traditional assets, particularly during market stress.}

Beyond their effectiveness as portfolio diversifiers, stablecoins have also been studied to detect their influence on other financial markets. Tether, for example, holds reserves—including money market funds—to back its issued stablecoins. Tether's buying or selling of these reserve assets can significantly impact the market prices of these reserves \cite{wu2023asset}. Additionally, stablecoins like TrueUSD, USD Coin, and Tether have a dynamic and bidirectional causal relationship with crude oil prices \cite{ghabri2022information}. This relationship likely reflects the interplay between the underlying asset, the US dollar, and oil prices.

Other studies have examined the interaction between stablecoins and cryptocurrencies. \citet{grobys2022tether} observed that jumps in Tether prices can predict a fall in Bitcoin prices, while \citet{alexander2022role} highlighted the role of Tether-margined contracts in transmitting volatility to Bitcoin prices. Traders often move assets between stablecoins and non-stable cryptocurrencies, depending on the market sentiment. As they get more bearish, they transfer their money from volatile cryptocurrencies to stablecoins, a phenomenon termed ``flight-to-cryptosafety''. Leveraged margined contracts amplify this relationship by enabling the transfer of larger sums. Several other authors, including \citet{paeng2024spillover} and \citet{gubareva2023stablecoins}, have similarly identified bidirectional causality and spillover effects between stablecoins, major cryptocurrencies, and traditional financial assets, highlighting the increasing connectivity of stablecoin markets within the broader financial system. However, this growing interdependence can reduce the effectiveness of stablecoins as portfolio diversifiers.

Conversely, some studies challenge the extent of these spillover effects. \citet{chen2022volatility} and \citet{kolodziejczyk2023stablecoins} reported minimal causality and spillover effects between stablecoins and other assets like cryptocurrencies or stocks, emphasizing stablecoins' relative stability. These contrasting findings are likely due to differences in methodologies, data granularity, sample periods, and model specifications. For instance, \citet{grobys2022tether} applied the methodology of \citet{barndorff2006econometrics} to high-frequency hourly price data spanning November 2018 to June 2021. In contrast, \citet{chen2022volatility} employed a combination of GARCH and VAR models on lower-frequency daily price data over a broader period from July 2014 to February 2022.

\insights{The presence of spillovers between volatile cryptocurrencies and stablecoins remains subject to ongoing debate.}

\begin{figure*}[htbp]
    \centering
    \includegraphics[width=\textwidth]{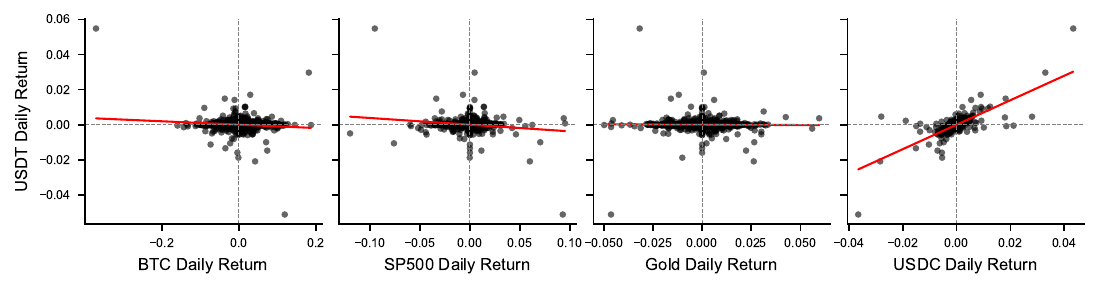}
    \caption{\journalColorText{Scatter plots of USDT daily returns against BTC, S\&P 500, USDC, and gold from 2020 to 2025. Each plot includes a fitted linear regression line. Data was sourced from \href{https://pypi.org/project/yfinance/}{Yahoo Finance API}.}}
    \label{fig:usdtscatterplots}
\end{figure*}

\journalColorText{Fig.~\ref{fig:usdtscatterplots} shows the negative or near-zero correlation between USDT and major financial assets like BTC, gold, and the S\&P 500, reinforcing its role as a diversifier or safe haven. In contrast, the strong positive correlation with USDC underscores the interconnectedness among stablecoins, suggesting potential channels for systemic risk.}

\textbf{Methodologies}
A range of econometric models has been employed to capture dependencies between the time series of stablecoin prices and other financial assets. These models reflect the need to account for characteristics such as volatility clustering, non-linear relationships, and behavior under extreme market conditions. These include GARCH, copula, quantile-based, dummy regression, vector autoregression, vector error correction, causality, and transfer entropy models.

Among these, GARCH models and their variants have been the most commonly employed. For example, GARCH models are employed in \cite{wang2020stablecoins,feng2024stablecoins,kliber2022looking} to analyze volatility dependencies. Asymmetric variants, such as EGARCH and GJR-GARCH, are sometimes chosen because negative price shocks can cause greater volatility in other assets compared to positive shocks \cite{kliber2022looking}, \cite{wang2020stablecoins}, \cite{feng2024stablecoins}. Additionally, DCC-GARCH models are applied in \cite{wang2020stablecoins} and \cite{feng2024stablecoins} to capture time-varying relationships between assets.

Compared to GARCH models, copula models better capture non-linear dependencies and have been employed by several authors \cite{belguith2024can, chen2022volatility}. Similarly, quantile-based models address non-linearity by analyzing comovements across different quantiles of the price distribution. Accounting for non-linearity is important because non-linearity has been documented in stablecoins' time series \cite{maiti2020dissecting, pernice2021stablecoin}.

Regression models with dummy variables were used in \cite{baur2021crypto} and \cite{gadi2022analyzing}. The authors regressed stablecoin price returns against dummy variables corresponding to significant declines in other assets, helping to classify stablecoins as safe havens.

Vector Error Correction (VEC) and Vector Autoregression (VAR) models were employed in \cite{anisiuba2021analysis} and \cite{paeng2024spillover}. These models capture the lead-lag relationship and cointegration between stablecoins and other assets. \citet{gubareva2023stablecoins} utilized a causality test to analyze the predictive relationships between stablecoins and other financial assets across different quantiles. \citet{paeng2024spillover} used a VAR model with quantile Granger causality.

\citet{ghabri2022information} used transfer entropy to study the information flow between energy markets, cryptocurrencies, and stablecoins. An advantage of the transfer entropy model is that it is nonparametric, requiring fewer assumptions about the underlying data-generating process.

Note that one limitation of many papers in this section is that they use models (e.g., VAR, VEC, GARCH, regression) that do not take into account the specific behaviors of stablecoin time series which have been demonstrated by other researchers, including non-linearity \cite{pernice2021stablecoin}, regime-switching behavior \cite{maiti2020dissecting}, and heavy-tailed distributions \cite{levene2021hypothesis}.

\insights{The most commonly used econometric models (e.g., VAR, VEC, GARCH, regression) do not take into account the specific behaviors of stablecoin time series which have been demonstrated by other researchers, such as non-linearity.}

\textit{\textbf{(2) How Do Macroeconomic, Blockchain, Media, Social, Climate, and Regulatory Factors Interact With Stablecoins}}

Stablecoins are influenced by a variety of factors. Rising interest rates typically lower their market capitalizations as investors shift to government bonds. While stablecoins can contribute to economic growth, they can also act as destabilizing forces in the economy. There is a positive bidirectional relationship between stablecoin activity and blockchain transaction fees. Economic uncertainty and negative sentiment drive demand for stablecoins as safe-haven assets. Media reports also influence stablecoins, though the impact varies depending on the type of stablecoin and the context of the media coverage. Unlike energy-intensive cryptocurrencies, stablecoins are not significantly affected by climate factors. Additionally, negative government policies and regulatory uncertainties tend to reduce demand for stablecoins.

Studies have found that higher U.S. federal funds rates and Chinese interbank rates reduce stablecoin prices and their price volatility while increasing their trading volumes \cite{nguyen2022stablecoins}. The price decrease is likely due to a shift in capital from stablecoins to interest-bearing assets like bonds or savings accounts following an interest rate hike. However, it remains unclear why stablecoin price volatility decreases despite the drop in prices.

\citet{bojaj2022forecasting} explored the macroeconomic effects of stablecoin adoption in Montenegro. They found that while stablecoins positively impact GDP, they may struggle to maintain their peg during market crashes, potentially destabilizing the economy.

\citet{osman2024economic} reported that economic sentiment significantly impacts the cryptocurrency market, especially during periods of uncertainty (e.g., COVID-19). Negative economic sentiment can increase demand for stablecoins as safe assets. Media announcements have a heterogeneous impact on stablecoin transaction volumes. For example, Tether (USDT) is particularly influenced by media reports, with both positive and negative news leading to increases in trading volumes. Decentralized stablecoins like DAI are more sensitive to institutional announcements compared to centralized stablecoins such as USDC and USDT \cite{de2024dollar}. 

\citet{ante2024time} found a bidirectional causal relationship between stablecoin transaction fees and stablecoin transactions: high fees deter users from transacting using stablecoins, and increased demand for stablecoin transactions drives up transaction fees as users compete for the limited capacity on the network. 

Environmental variables, such as $CO_2$ emissions and temperature anomalies, significantly predict the returns of volatile cryptocurrencies like Bitcoin and Ethereum. However, these environmental factors do not predict the returns of stablecoins like Tether (USDT) and TrueUSD. This indicates that investors do not associate stablecoin returns with environmental factors due to their non-energy-intensive nature \cite{clark2023cryptocurrency}. 

\citet{ayadi2023directional} explored the impact of CBDCs on stablecoins, finding that the CBDC Uncertainty Index is negatively related to stablecoin returns. High uncertainty around CBDC developments can decrease stablecoin valuations, reflecting investors' concerns about competition or regulatory uncertainties.

Also, governmental regulatory policies can significantly affect stablecoins. For instance, the Chinese government's regulatory actions in May 2021 had a markedly negative impact on the cryptocurrency market, including stablecoins \cite{su2022analysis}. Chinese regulators announced that cryptocurrencies could not be used as currency and prohibited many cryptocurrency-related services.

\insights{Stablecoin markets are influenced by a variety of factors, including interest rates, economic sentiment and uncertainty, and regulatory policies.}

\insights{Stablecoin activity has been found to influence macroeconomic factors, 
transaction fees on blockchains, and the traditional finance market (e.g., Tether’s reserve asset purchases impacting money markets).}

\textbf{Methodologies:} To address the research question, "How do macroeconomic, media, social, environmental, or regulatory factors interact with stablecoins?" researchers have employed a range of econometric and statistical methods, including GARCH, Granger causality, VAR, and regression models. For instance, \citet{nguyen2022stablecoins} used GARCH and EGARCH models to analyze the impact of interest rates on stablecoin prices, selecting EGARCH to account for the asymmetric effects of interest rate increases and decreases. Similarly, \citet{ante2024time} applied time-varying Granger causality to investigate the bidirectional relationship between Ethereum transaction fees and economic activity in stablecoins, highlighting the robustness of time-varying methods in capturing relationships that evolve with market conditions.

\citet{bojaj2022forecasting} utilized a Bayesian VAR model to forecast the macroeconomic effects of stablecoin shocks on the Montenegrin GDP. However, the lack of sufficient control variables in their model suggests the possibility of confounding factors influencing both stablecoins and GDP. 

Another group of authors, including \citet{de2024dollar} and \citet{su2022analysis}, adopted event study approaches, measuring abnormal price and trading volume changes surrounding media announcements or regulatory developments. 

Predictive regression models have also been employed. For example, \citet{clark2023cryptocurrency} regressed stablecoin returns on variables related to CO2 emissions and temperature anomalies. 

Meanwhile, \citet{ayadi2023directional} used a Cross-Quantilogram model to analyze non-linear correlations between CBDC-uncertainty and CBDC-attention indices and changes in stablecoin prices and trading volumes across different quantiles.

\textit{\textbf{\journalColorText{(3) How Do Stablecoin Issuances/Transfers Impact Non-Stable Cryptocurrencies}}}


\journalColorText{Several studies provide evidence that stablecoin issuances and transfers can impact non-stable cryptocurrencies, particularly Bitcoin, in the short term \cite{ante2021influence, ante2021impact, saggu2022intraday, griffin2020bitcoin}. These impacts manifest as positive abnormal trading volumes and returns surrounding issuance or transfer events. Stablecoin issuances may provide liquidity to the cryptocurrency markets, enabling more trading activity and potentially driving up prices \cite{wei2018impact, ante2021influence}. Investor sentiment regarding stablecoin issuance may also play a role in this effect \cite{saggu2022intraday}.}

\begin{figure}[htbp]
    \centering
    \includegraphics[width=\linewidth]{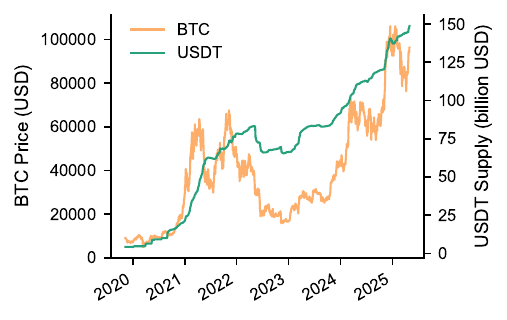}
    \caption{\journalColorText{Daily BTC price and USDT circulating supply (in billions of USD) from 2020 to 2025. Data was sourced from \href{https://developers.coindesk.com/}{CoinDesk API}.}}
    \label{fig:btcusdt}
\end{figure}

\journalColorText{However, the nature of these impacts remains debated. For example, \citet{griffin2020bitcoin} suggested that unbacked Tether issuances may have been used to manipulate Bitcoin prices, indicating a direct and intentional impact on non-stable cryptocurrencies. On the other hand, \citet{wei2018impact} found no causal impact of Tether issuances on Bitcoin returns.}

\journalColorText{Adding further nuance, some studies suggest that stablecoin issuances are a reaction to increased demand for cryptocurrencies during bullish markets rather than a cause of price increases \cite{kristoufek2021tethered, kristoufek2022role}. This perspective implies that stablecoins may not have a significant causative impact on non-stable cryptocurrency prices.}

\journalColorText{The discrepancies in findings may stem from differences in data and methodologies used. For instance, \citet{griffin2020bitcoin} employ clustering and regression models on highly granular intraday transaction-level blockchain data, while \citet{wei2018impact} relies on Granger Causality and VAR models using daily price data.}


\insights{\journalColorText{Stablecoin issuances tend to be correlated with higher trading volumes and abnormal returns of non-stable cryptocurrencies.}}

\journalColorText{The co-movement between stablecoin issuance and non-stable cryptocurrency prices is visually apparent in Fig.~\ref{fig:btcusdt}, where increases in USDT supply often align with rising Bitcoin prices. However, this visual correlation does not imply causation and may, for instance, reflect demand-driven issuance during bullish market periods \cite{kristoufek2022role}.}


\journalColorText{\textbf{Methodologies: }The research question, "How Do Stablecoin Issuances/Transfers Impact Non-Stable Cryptocurrencies" is addressed using various methodologies. \citet{griffin2020bitcoin} analyze blockchain transactions. They use clustering algorithms and regression models. The clustering algorithms are used to group wallets into specific entities, such as different exchanges. The flow of USDT and Bitcoin between these entities is analyzed. The authors use regression models with Bitcoin price changes as the target variable, and tether flows as the explanatory.}

\journalColorText{Several other authors regress Bitcoin price changes (abnormal returns) against variables related to USDT transfers or supply \cite{ante2021impact, saggu2022intraday, ante2021influence, wei2018impact}. Abnormal returns are calculated based on an event-study methodology. In other words, they are calculated as the difference between returns close to USDT issuance events and the average of returns over a specified time period. \citet{ante2021influence} used a t-test to determine whether Bitcoin's abnormal returns around USDT issuance events were statistically different from zero. In addition, several papers used vector autoregression models (VAR) to study the relationship between USDT issuances and Bitcoin returns and trading volumes \cite{wei2018impact, kristoufek2021tethered, kristoufek2022role}.}

\textit{\textbf{\journalColorText{(4) How Do External Crises Impact Stablecoins?}}}

\journalColorText{The main results of this section are the following: the impact of crises on stablecoins depends on the nature of the crisis; for instance, stablecoins holding money at failing banks are negatively affected, while stablecoins with more trust can increase in capitalization during crises; nevertheless, there is evidence that investors prefer traditional safe assets like government bonds during crises.}

\journalColorText{During periods of market distress, investors often move capital into stablecoins to preserve value, resulting in heightened demand and increased capitalization for these assets \cite{kyriazis2019cryptocurrencies, galati2024market}.}

\journalColorText{Nevertheless, external crises can spread volatility from other cryptocurrencies to stablecoins, impacting their stability and diminishing their role as safe havens. For instance, following the collapse of the SVB bank, increased interconnectedness between stablecoins and volatile cryptocurrencies became evident, with one of the largest stablecoins (USDC) losing around 13\% of its value. This decline was likely due to the stablecoin holding reserves with the failed bank. Conversely, some crises can have positive effects on stablecoins. For instance, during the collapse of the FTX digital currency exchange, investors exhibited flight-to-safety behavior by transferring funds to certain stablecoins. Not all stablecoins are perceived equally: in times of crisis, investors tend to generally favor those with stronger backing or better reputations over lesser-known or less authoritative options \cite{galati2024silicon, oefele2024flight}.}

\begin{figure}[htbp]
    \centering
    \includegraphics[width=\columnwidth]{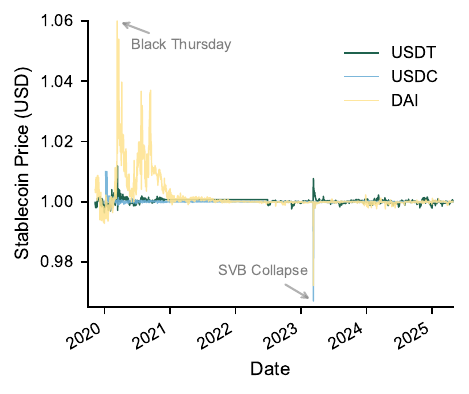}
    \caption{\journalColorText{Time series of major stablecoin prices, with two crises annotated: Black Thursday (COVID-19 crash), and Silicon Valley Bank collapse.}}
    \label{fig:stablecoin_crises}
\end{figure}

\journalColorText{The FTX crisis also caused market discrepancies between exchanges (different prices of the same stablecoin) \cite{galati2024market}. Exchanges can generally keep asset prices aligned through arbitrage, where traders exploit price differences by buying low on one exchange and selling high on another. However, this process is not instantaneous due to factors such as blockchain transaction speeds and withdrawal/deposit limitations on exchanges. When price changes are too rapid, arbitrage may not be able to keep up, leading to temporary price discrepancies. Crises can also trigger a sudden surge in fund withdrawals, which some exchanges may be unable to meet immediately due to operational or liquidity constraints. Additionally, some exchanges are riskier than others, and investors may price in this risk for assets listed on higher-risk exchanges.}

\insights{\journalColorText{As investors get more bearish, they transfer their money from volatile cryptocurrencies to stablecoins, a phenomenon termed ``flight-to-cryptosafety''.}}

\journalColorText{The impact on stablecoins depends on the characteristics of the crisis and the stablecoins involved. This was evident during the SVB banking crisis. Fig.~\ref{fig:stablecoin_crises} shows that during this crisis, USDC---which held a significant amount of reserves with the failed bank---suffered a significant depeg. DAI followed suit due to its reliance on USDC as collateral. In contrast, USDT remained stable, shielded by its lack of exposure to SVB; demand for USDT even increased as investors shifted from USDC and DAI. USDC and consequently DAI recovered quickly following the U.S. regulators' announcement that Silicon Valley Bank depositors would be protected \cite{oefele2024flight}.} 

\journalColorText{The Black Thursday crypto-market crash in March 2020 offers a different example. During a sharp decrease in demand for risky assets, driven by negative investor sentiment related to COVID-19 (Bitcoin prices, for example, fell over 40\% in one day), investors quickly moved funds into stablecoins, seeking safety \cite{redman2023bitcoins}. This sudden increase in demand overwhelmed the minting and arbitrage mechanisms of several major stablecoins, causing them to temporarily lose their peg and trade above \$1 (visible in Fig.~\ref{fig:stablecoin_crises}). DAI experienced a particularly large de-peg during this time, due to problems specific its oracle price feeds and liquidation auction system \cite{eichholz2020what, klagesmundt2019instability}.}

\insights{\journalColorText{In severe crises, investors may prefer traditional safe assets (government bonds) over stablecoins, leading to outflows from the stablecoin market.}}

\journalColorText{\textbf{Methodologies: }The research question, "How Do External Crises Impact Stablecoins?" is addressed by four papers using different methodologies, all studying the price or volume of stablecoins during a specific period surrounding a crisis. \citet{oefele2024flight} used descriptive statistics to analyze the total net asset changes of 11 stablecoins and 223 money market mutual funds during the March 2023 Silicon Valley Bank crisis. They compare the inflows and outflows of these assets before and after the crisis. \citet{kyriazis2019cryptocurrencies} compared the Amihud's illiquidity ratio of 846 cryptocurrencies during the bearish market phase from April 2018 to January 2019. Amihud's illiquidity ratio measures how much an asset's price moves per unit of trading volume. Based on this measure, they found the most liquid assets during the stress phase. A criticism of the methodology in \cite{kyriazis2019cryptocurrencies} is that it does not compare liquidity ratios during the stress phase with those outside of the stress phase.
\citet{galati2024market} used t-tests to check whether the liquidity of six major cryptocurrencies decreased after the crypto-exchange FTX's withdrawal freeze. \citet{galati2024silicon} used a BEKK-GARCH model to study volatility spillovers between Bitcoin and five major stablecoins following the March 2023 banking crisis. They also calculate abnormal returns and trading volumes following the crisis.}



\textbf{Research Gaps:} This category of research focuses on the comovement of stablecoins with other assets or factors. However, only one paper discusses the comovement of different stablecoins with each other \cite{thanh2023stabilities}. This study is limited to five prominent stablecoins, using a relatively short sample period of 1.5 years, and relies solely on daily pricing data. In addition, the investigation does not consider non-linear relationships or regime changes. Future research should encompass a broader selection of stablecoins and use more sophisticated methodologies, possibly incorporating high-frequency data. 

There also remains a debate regarding whether the correlation identified between stablecoins (particularly Tether) and cryptocurrencies (particularly Bitcoin) is a causative one. \citet{griffin2020bitcoin} argued, for instance, that there is a causal relationship as Tether issuances manipulate Bitcoin prices upwards, while other authors (e.g., \citet{wei2018impact}) argued against this---we provide a more detailed analysis of the reasons behind these conflicting results in Section~\ref{sec:discussion}.

Only one paper studies the impact of interest rates on stablecoins \cite{nguyen2022stablecoins}. This study is not conclusive. It studies a relatively short sample of one year, which is not enough given the infrequent and longer-term nature of interest-rate changes. 

Moreover, the relationship between stablecoins and other unstudied factors is necessary. For example, there might be a significant relationship between activity on some DeFi platforms and stablecoins, such as Aave (a platform offering stablecoin lending), Curve Finance (a stablecoin trading decentralized exchange), and Yearn Finance (a platform that optimizes yield from stablecoins). There is also a lack of studies on the impact of announcements by stablecoin providers on their own stablecoins. The relationship between stablecoins and emerging markets, e.g., Brazil, Nigeria, Turkey, Indonesia, and India, also necessitates more attention since some stablecoins are heavily used in those countries~\cite{sandor2024stablecoins}.

Overall, 51 different stablecoins are studied in this area. The most studied stablecoins are USDT (52 papers), USDC (34 papers), DAI (20 papers), TUSD (16 papers), BUSD (14 papers), USDP (13 papers), GUSD (11 papers), and HUSD (5 papers). 
Despite the relatively high number of stablecoins analyzed in existing work, there exists a significant opportunity for conducting research on stablecoin projects that have received limited to no attention from academia. This can give us insight into the stablecoins that have different characteristics from the ones considered in existing studies. One example of a different stablecoin, which has been neglected by academic investigations so far, is RAI. Unlike the majority of stablecoins that are pegged to a fixed price (usually one USD), RAI uses a Proportional-Integral-Derivative (PID) mechanism to dynamically adjust the peg value.


\subsection{\journalColorText{Stablecoin Usage and Adoption}}


\journalColorText{\textit{\textbf{(5) What Are The Economic, Financial, And Policy Impacts Of Adopting Stablecoins?}}}

\journalColorText{The main results of this section are the following: issuance of stablecoins shifts deposits from commercial banks to the stablecoin issuer, destabilizing banks; CBDCs can lead to synchronized interest rates across countries, limiting monetary policy autonomy; the presence of stablecoins reduces the effectiveness of central banks' open-market operations.}

\journalColorText{\citet{castren2022digital} argued that the introduction of central bank digital currencies (CBDC) creates a funding gap for commercial banks as deposits shift from these banks to the central bank. Commercial banks would be forced to fill this gap through various methods. The authors show that none of the methods (even with the cooperation of central banks) would be fully effective in offsetting the higher costs and potential instability that would ripple through the economy. The authors show how the introduction of privately owned stablecoins would similarly alter the financial network, shifting deposits from commercial banks to the issuing entity and creating a funding gap for the banks. \citet{karau2023central} used a theoretical two-country model to show how the issuance of a CBDC would increase interest-rate synchronization between countries, leading to less monetary policy autonomy for them.} 

\begin{figure}[htbp]
    \centering
    \includegraphics[width=0.7\columnwidth]{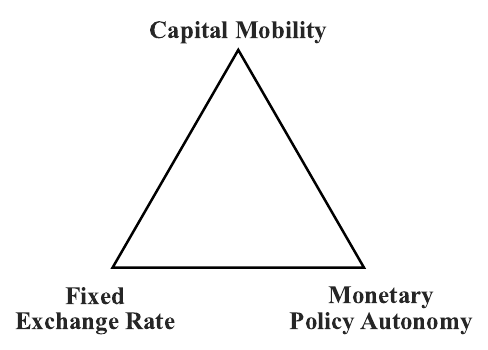}
    \caption{\journalColorText{The Impossible Trinity, or Trilemma, in international economics. It states that a country can simultaneously achieve at most two of the following three policy objectives at a given time---capital mobility, fixed exchange rate, and monetary policy autonomy.}}
    \label{fig:impossible_trinity}
\end{figure}


\journalColorText{\citet{park2023stablecoins} argued that the introduction of privately issued stablecoins to a financial system limits the effectiveness of open-market operations by central banks. For instance, central banks often issue bonds to decrease the money supply (since investors exchange money for bonds). However, in the existence of stablecoin issuers, the issuers can purchase government bonds and then issue currency backed by these bonds, increasing the money supply again. However, the authors show that central banks can mitigate against this by using other monetary policy tools instead, such as raising interest rates on reserves held at central banks.}

\journalColorText{The introduction of stablecoins could challenge traditional economic assumptions, such as the Mundell-Fleming monetary trilemma \cite{aizenman2019modern}, illustrated in Fig.~\ref{fig:impossible_trinity}. In this case, the accessibility and non-sovereign nature of stablecoins undermine the ability of governments to enforce capital controls, making the trilemma more difficult to manage.}

\journalColorText{\textbf{Methodologies: }The three papers addressing the research question, "What Are The Economic, Financial, And Policy Impacts Of Adopting Stablecoins?" do so using theoretical modeling.}

\journalColorText{\citet{castren2022digital} modeled the financial system using a network with 10 connected nodes, each representing a different entity (e.g., households, non-financial corporations, banks, the central bank). The paper simulates a shift in deposits from commercial banks (MFIs) to the central bank (in the case of CBDC introduction) or investment funds (in the case of stablecoin introduction). The authors then show how this shift can impact the different nodes and links of the financial network. The authors use real data from the Euro Area Accounts (EAA). This data allows them to identify the different entities in the network and the centrality of each entity based on the assets and liabilities held in relation to other entities. Using real data, they simulate the shock that would be caused by CBDC or stablecoin issuance under different scenarios.}

\journalColorText{\citet{karau2023central} developed a two-country model to analyze the impact of central bank digital currencies (CBDCs) on interest rates and monetary policy. Their mathematical framework integrates equations for interest rate parity, liquidity services, and money demand to demonstrate how the introduction of CBDCs drives currency competition, compelling countries to synchronize their interest rates.}

\journalColorText{The authors of \cite{park2023stablecoins} modify the established Lagos and Wright model \cite{lagos2005unified} to analyze stablecoins' impact on monetary policy effectiveness. The model includes buyers of goods, sellers of goods, and money issuers. Time is divided into periods; in each period, participants can trade money, bonds, and reserves. A key parameter of the model is the pledgeability parameter, which refers to the fraction of the collateral that stablecoin holders expect the issuer to actually pay back when redeeming. The authors argue that this parameter is less than 1; this could be because stablecoin issuers might not be perfectly honest about the collateral they own or because some stablecoins are not fully collateralized. The mathematical model used reflects the utility of the agents (buyers, sellers, and issuers) and the assets involved (money, reserves, and government bonds). The equilibrium of this system occurs when the stablecoin issuers’ profit from issuing money in the form of stablecoins is zero.}


\journalColorText{\textit{\textbf{(6) What Factors Influence The Adoption Of Stablecoins?}}}

\journalColorText{The results of this section are the following: factors that explain stablecoin success include increased stability, the presence of trustworthy reserves, high liquidity and accessibility, early mover advantages and network effects, and cross-border payment efficiency; factors that hinder their success include regulatory uncertainty, technical risks such as smart contract failures, lack of user technical knowledge, and privacy concerns.}

\journalColorText{The literature discusses several factors that explain the adoption of stablecoins. Expectedly, their stability motivates their adoption, as it makes them more attractive as a medium of exchange and reduces risks \cite{au2024characteristics, wlosik2023win}. The presence of reserves backing stablecoins increases user confidence, which is important for adoption \cite{hsu2023revealing}. Factors related to liquidity and ease of transaction, including being listed on many cryptocurrency exchanges, available in a wide variety of trading pairs, and integrated into multiple blockchain platforms, also boost adoption \cite{son2022consumer, au2024characteristics}. Early mover advantage and network effects are also significant predictors of stablecoin success or adoption \cite{hsu2023revealing}. Early mover advantage allows stablecoins to gain a larger market share and integration into the cryptocurrency ecosystem, making them more valuable and trusted. Another factor motivating the adoption of stablecoins is improved cross-border payment efficiency, making them attractive for remittances \cite{singh2023quest}.}

\insights{\journalColorText{Regulatory uncertainty and compliance, technical risks, and lack of skills significantly limit retail and institutional adoption of stablecoins.}}

\journalColorText{The lack of clear regulations and the varying legal classifications of stablecoins across jurisdictions are substantial barriers to their adoption. Issues related to money laundering, terrorist financing, and consumer protection are among the most significant regulatory risks related to stablecoins \cite{sood2023identification, adams2023investigating}. Moreover, governments and central banks are concerned about stablecoins eroding monetary sovereignty, limiting their support to stablecoins and leading to the exploration of CBDCs as a response \cite{singh2023quest}. Technical risks, particularly related to blockchain technology and smart contract failures, are major concerns for stablecoin adoption. Stablecoins are vulnerable to operational failures, oracle risks, and immature decentralized technologies \cite{sood2023identification}. The surveyed experts mentioned the following challenges hindering stablecoin adoption in the finance industry of South Africa: lack of blockchain infrastructure support, scalability, energy usage, privacy and security, lack of organizational skills, regulatory uncertainty, negative public perception of cryptocurrency \cite{adams2023investigating}. The technical know-how of users also plays a role. While experienced users can appreciate stablecoins for their price stability, novice users may find the concept of stablecoins confusing \cite{au2024can}. Consumer concerns over privacy and data protection also act as deterrents to wider adoption \cite{sood2023identification}.}

\journalColorText{\textbf{Methodologies: }To understand what factors influence stablecoin adoption, researchers used interview, netnography, experimental, and theoretical-modeling methodologies. Interviews with industry experts were conducted in \cite{adams2023investigating, singh2023quest, sood2023identification}. To facilitate the interviews, factors can also be identified using a systematic literature review \cite{sood2023identification}. \citet{au2024characteristics} and \citet{hsu2023revealing} used netnography, analyzing discussions of online users, to identify frequently mentioned factors that motivate stablecoin adoption. \citet{au2024can} employed an experimental design where 225 participants answered a series of questions before and after being exposed to an advertisement about stablecoins. \citet{baughman2023global} developed a theoretical model that includes a basket-based stablecoin of the currencies of two countries. Through this model, they analyze the demand for this stablecoin, reaching an equilibrium of limited demand. \citet{son2022consumer} developed a theoretical model that analyzes how users maximize their utility by allocating their funds across different liquid assets: cash, stablecoins, CBDCs, and deposits. The model includes variables related to these assets, including whether it is recognized as legal tender, anonymous, digital, and their interest rate.}

\journalColorText{\textit{\textbf{(7) What Are Current Use Cases Of Stablecoins?}}}

\journalColorText{The main results of this section are the following: use-cases of stablecoins include as collateral on DeFi lending platforms, payments and remittances, speculative trading, and as rewards for crowdsourcing tasks; the presence of leverage in stablecoin use-cases and the interconnectedness of stablecoin usage with the broader DeFi ecosystem presents a systemic risk.}

\insights{\journalColorText{Stablecoins are used for payments and also serve as key assets in lending, borrowing, and leveraged trading.}}

\journalColorText{\citet{darlin2022debt} highlighted that stablecoins are used in leveraged lending. Stablecoins are used in complex financial strategies where users use debt obtained from one DeFi platform as collateral to obtain more debt from another platform. This leveraged lending can magnify returns from arbitrage, speculation, or yield farming. \citet{wang2022speculative} also identified the use of stablecoins in leveraged speculation using similar debt-financed-collateral strategies. \citet{tovanich2023contagion} found similar debt-financed-collateral usage of stablecoins on the popular Compound DeFi lending protocol. \citet{rosa2021tether} argued that patterns in the stablecoin transaction network are evidence that it is being used in speculative trading, fueling speculative bubbles. \citet{ante2023profiling} highlighted the important role stablecoins serve as a payment method in Turkey. \citet{meng2023cryptocurrency} highlighted the use of stablecoins in task crowdsourcing platforms as payment and concluded that tasks paid in stablecoins are more popular than those paid by unstable cryptocurrencies.} \journalColorText{Several of the papers also highlight the high systemic risk arising from the usage of stablecoins \cite{darlin2022debt, tovanich2023contagion, wang2022speculative}. Systemic risk is reflected in the potential effects of stablecoin failures, which can heavily impact various types of DeFi platforms, such as lending platforms, decentralized exchanges, and yield farming platforms. This risk increases as stablecoins become more widely adopted, and their adoption is increasing rapidly \cite{coingecko2024}.} 

\journalColorText{\textbf{Methodologies: }To find current uses of stablecoins, several methodologies have been used. \citet{meng2023cryptocurrency} employed regression analysis on crowdsourcing platform data to study the impact of stablecoin rewards on task participation. \citet{tovanich2023contagion} constructed a financial network from the DeFi platform Compound's on-chain data. This network is then analyzed to understand how stablecoins are used. Compound is a DeFi platform that allows users to lend and borrow various cryptocurrencies, including stablecoins. \citet{rosa2021tether} also constructed a network of USDT transaction data to understand USDT usage. \citet{darlin2022debt} analyzed on-chain transaction data from various DeFi protocols to identify debt-financed collateral usage. \citet{ante2023profiling} surveyed Turkish cryptocurrency owners to understand their stablecoin usage patterns. \citet{khan2023way} implemented theoretical models using computational simulation of cross-border payments with stablecoins, conducting experiments to understand transaction speed and costs of different payment systems. \citet{wang2022speculative} developed a formal theoretical model of on-chain leveraged trading using stablecoins.}


\journalColorText{\textit{\textbf{(8) What Are The Potential Use Cases For Stablecoins Across Different Industries?}}}

\journalColorText{The main results of this section are the following: stablecoins can enable micropayment and streamline transactions for telecom companies; stablecoins can enable transparent returns, loyalty programs, and supply-chain financing in the retail industry; stablecoins can improve payment for international travelers; stablecoins can facilitate timely payments, escrow, and installment processes in the construction industry.}

\journalColorText{Four papers discuss potential use cases of stablecoins for specific industries \cite{zhang2024blockchain, ayuba2022conceptual, bhat2022telco, manahov2024stablecoins}. \citet{bhat2022telco} argued that telecom companies could use stablecoins to facilitate payments, particularly micropayments to IoT devices. Stablecoins can also reduce transaction costs and speeds for payments to telecom operators and payments by users. Their ability to be programmed using smart contracts also makes them advantageous for use cases such as paying open-source project developers and getting payments from users. \citet{zhang2024blockchain} argued that stablecoins are suitable for retail payments, particularly with product returns and loyalty programs. It can also be used for supply chain financing. \citet{manahov2024stablecoins} argued that stablecoins could make payments in the travel industry more convenient by making international payments more efficient. \citet{ayuba2022conceptual} argued that stablecoins can be useful for payments in the construction industry. They can help increase trust that the clients will pay on time, facilitate escrow processes, and streamline construction progress installments. Stablecoins also open up international financing opportunities for real estate developers, particularly for regions with limited access to traditional financing.}

\journalColorText{\textbf{Methodologies: }To argue for potential use cases of stablecoins in specific industries, the four papers utilize a mix of conceptual approach, mathematical modeling, literature review, and semi-structured interviews. Some authors employed a purely conceptual approach, with no empirical or formal theoretical analysis \cite{bhat2022telco, manahov2024stablecoins}. \citet{zhang2024blockchain} constructed a formal mathematical model of retailer and consumer utility. The model involves a single retailer and a mass of consumers. The retailer chooses between stablecoin payments and traditional payment methods (credit card). The parameters of the model include production cost, credit card processing fees, salvage value of returned products, price of the product, the amount refunded to consumers, cost of information search by consumers, the value consumers give the product, etc. Through the equations of this model, the authors show the value of using stablecoins in retail payments for retailers and consumers. \citet{ayuba2022conceptual} conducted semi-structured interviews with industry experts to understand the use-case of stablecoins in the construction industry.}

\journalColorText{\textbf{Research Gaps: }Researchers have identified several future research directions to address current gaps related to the topic of digital currency usage and adoption. Future research could explore how the design of a stablecoin affects its usage and impacts \cite{castren2022digital}. Additionally, studies can investigate how adopting CBDCs or stablecoins impacts international trade and financial stability \cite{singh2023quest, baughman2023global, au2024can}. There is also a need for more theoretical macroeconomic models that integrate CBDCs and stablecoins, which would allow economists to simulate their broader impacts on inflation, banking, and monetary policy \cite{sood2023identification, darlin2022debt}. Some researchers have highlighted the ongoing debate regarding the impacts of digital currencies, with arguments both in favor of their potential benefits and their significant risks. \citet{son2022consumer} offered theoretical models on how consumers and producers will behave with the introduction of stablecoins but emphasized the need for empirical research to validate these findings. Some authors surveyed users and experts in Turkey and South Africa \cite{bhat2022telco, manahov2024stablecoins}; further research is necessary to analyze the usage and perceptions of digital currencies in more countries.}

\journalColorText{The three papers addressing the research question, "What are the economic, financial, and policy impacts of adopting stablecoins?" primarily rely on theoretical modeling \cite{castren2022digital, karau2023central, park2023stablecoins}. This highlights a gap in the literature: the lack of empirical research on this topic. For example, empirically analyzing the shift in the financial network structure following the adoption of a stablecoin or CBDC in an economy would provide valuable insights.} 

\journalColorText{Additionally, the theoretical models discussed are built upon a set of assumptions, variables, and model structures. Employing differing combinations of these factors could yield varying results, leaving room for future theoretical model development. For instance, the two-country model presented in \cite{karau2023central} could be expanded to incorporate multiple countries. \citet{castren2022digital} considered only the immediate effects of stablecoin introduction on the financial network using a one-period model. Future research could extend the analysis to multiple periods to study the long-term impacts and higher-order adjustments. Additionally, the impact of incorporating factors such as varying risk preferences among different entities in the network, or the potential for strategic cooperation and government intervention, could be explored.}



\subsection{Stablecoin Stability Analysis}

\textit{\textbf{(9) How Stable Are Stablecoins In General Market Conditions?}}

Stablecoins exhibit significantly lower volatility compared to non-stable cryptocurrencies but are still not as stable as the fiat currencies they are pegged to. Their stability is influenced by market conditions and their interconnectedness with non-stable cryptocurrencies, which can increase volatility. Some stablecoins are more stable than others. Generally, algorithmic stablecoins are less stable than collateralized ones.

The findings from seven papers suggest that stablecoins, despite their intended stability, exhibit varying degrees of instability in general market conditions. \citet{grobys2021stability} highlighted that stablecoins display unpredictable volatility. \citet{zhao2021understand} found that there are scenarios where the stability mechanisms of stablecoins fail, causing volatility. \citet{hoang2021stable} emphasized that stablecoins are relatively stable compared to Bitcoin but less so than fiat currencies, being influenced by Bitcoin's volatility. \citet{duan2023instability} revealed that many stablecoins deviate from their \$1 peg, although some, like BUSD, correct these deviations more effectively than decentralized stablecoins like DAI. \citet{fernandez2024extremely} discussed how systemic risks make stablecoins experience volatility, but also noted stablecoins' mean-reversion characteristics. \citet{hairudin2024isotropy} reinforced this by demonstrating that stablecoins, especially USDT, exhibit mean reversion and lower volatility compared to cryptocurrencies like Bitcoin and Ethereum.

Overall, while stablecoins tend to be less volatile than other cryptocurrencies, they are far from perfectly stable, with their volatility influenced by market conditions, systemic risks, and design mechanisms. However, being more volatile than intended does not make them valueless; the fact that they have significantly lower volatility than non-stable cryptocurrencies makes them of practical value to the market. 

\insights{Stablecoins exhibit significantly lower volatility compared to non-stable cryptocurrencies but are still not as stable as the fiat currencies they are pegged to.}

\insights{Some stablecoins are more stable than others. Generally, algorithmic stablecoins are less stable than collateralized ones.}

\textbf{Methodologies:} To assess the stability of stablecoins in general market conditions, researchers used descriptive statistics, power law analysis, hypothesis testing, stationarity analysis, formal theoretical modeling, and wavelet analysis. Descriptive statistics, such as standard deviation, mean, and range, are commonly used; for a perfectly stable stablecoin, all these measures would equal zero. Power law analysis has been applied to model the returns of stablecoins, with a power law exponent below three indicating that extreme events contribute to persistent volatility \cite{grobys2021stability}. It is important to note that the power law focuses on the tail behavior of distributions, modeling returns above a certain threshold. For instance, \citet{grobys2021stability} identified a volatility threshold of 0.175, above which the power law signals instability—though this does not imply instability under normal conditions. \citet{hoang2021stable} applied statistical hypothesis tests like the Chi-square test to determine whether the standard deviation of the stablecoin price series is significantly greater than zero. Additionally, \citet{duan2023instability} tested for stationarity of stablecoin returns around zero, where a lack of stationarity would indicate regime changes and deviations from zero average returns.

Wavelet analysis, used in \cite{hairudin2024isotropy}, allows researchers to assess the volatility of a time series across a range of frequencies or time scales. This is useful because a time series might be, for instance, not volatile over the long term but show volatility over short-term timescales, or vice versa.

\citet{charoenwong2023computer} used several computer science theories (undecidability and the Halting Problem) to analyze algorithmic mechanisms in stablecoin stability, transforming economic stability questions into computational abstractions. 

\citet{zhao2021understand} developed a formal modeling framework utilizing six classes of timed automata to represent stablecoin expansion, contraction, traders, and exchanges. Using the Uppaal model checker, they simulated all possible state transitions of the system, uncovering scenarios where expansion or contraction mechanisms fail due to undesirable trader behaviors. Empirical validation of these theoretical findings with blockchain transaction data revealed that trader behaviors contributed to stability failures in stablecoins such as Basis Cash and Ampleforth.

\textit{\textbf{(10) How Stable Are Stablecoins During Periods Of Market Turbulence Or Crisis?}}

Stablecoins, while generally more stable than non-stable cryptocurrencies, are not immune to volatility during market crises. For example, some stablecoins experienced heightened volatility during the COVID-19 pandemic. The Terra-Luna collapse triggered increased volatility in several stablecoins but drove higher demand and stability in others. Similarly, the Silicon Valley Bank crisis caused significant volatility in USDC, while the collapse of FTX led to fluctuations in both USDC and USDT.

During the COVID-19 pandemic, gold-backed stablecoins exhibited heightened volatility, with some surpassing Bitcoin in instability \cite{jalan2021shiny}. Unlike real gold, many gold-backed stablecoins failed to act as safe havens. Nevertheless, some gold-backed stablecoins performed better than others (e.g., PAX Gold). Other custodian-based USD-pegged stablecoins like Tether (USDT) and USD Coin (USDC) maintained low volatility and grew in market capitalization, while on-chain collateralized stablecoins like DAI experienced higher instability \cite{jeger2020analysis}.

\begin{figure}[htbp]
     \centering
     \includegraphics[width=0.48\textwidth]{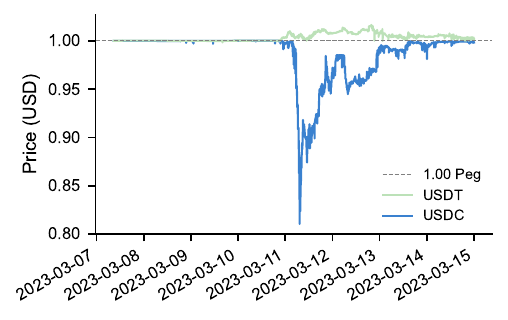} 
     \caption{\journalColorText{Minute-by-minute Prices of USDT and USDC during the SVB Crisis. Data was sourced from \href{https://developers.coindesk.com/}{CoinDesk API}.}}
     \label{fig:usdc-svb-crisis}
\end{figure}

The Terra-Luna collapse caused volatility across multiple stablecoins, yet collateral-backed stablecoins experienced increased demand and maintained relative stability \cite{de2023intelligent}. \journalColorText{The Silicon Valley Bank crisis caused significant volatility in USDC because a portion of its reserves were held at the failed bank, whereas USDT, which did not have direct exposure, remained stable during the same period (see Fig.~\ref{fig:usdc-svb-crisis}). USDC quickly recovered as U.S. regulators announced that all depositors at Silicon Valley Bank would be fully protected, even above the insurance limits \cite{oefele2024flight}.} The FTX collapse led to significant fluctuations in USDC and USDT, with both experiencing greater instability than Bitcoin and Ethereum \cite{hairudin2024fractal}.
These findings indicate that while stablecoins demonstrate some resilience, they are susceptible to unexpected volatility and external shocks during crises, with the impact varying based on the stablecoin's characteristics and its relationship to the crisis.

\insights{Stablecoins, while generally more stable than non-stable cryptocurrencies, are not immune to volatility during market crises. The extent of this volatility depends on both the characteristics of the stablecoin and the nature of the crisis.}

\textbf{Methodologies:} Methodologies similar to those employed in addressing the previous research question may also be applied here but with an adaptation to an event-study framework in order to examine volatility during specific events or crises. For instance, \citet{de2023intelligent} measured abnormal returns during the UST collapse. Abnormal returns refer to the difference between realized and expected returns. \citet{hairudin2024fractal} used wavelet analysis during several crises, including the FTX collapse and the COVID pandemic. 
\citet{jalan2021shiny} calculated realized variance (the square of the price changes) during the COVID-19 pandemic. In addition to the methodologies mentioned in the previous research question, the quantile-root test has been used in \cite{jalan2021shiny}.
Quantile Unit Root tests are used to determine if a time series is stationary or not by examining the stationarity of different quantiles of its distribution. This is useful because extreme market conditions will fall in upper or lower quantiles. A non-stationary time series would be unstable.




\textbf{Research Gaps:}
Studies on stablecoin stability focus on collateralized designs, such as USDT, USDC, and DAI, while the stability of purely algorithmic stablecoins remains understudied. Greater comparative analysis is needed to compare the stability of stablecoins with different characteristics (e.g., collateralized vs. algorithmic, varying collateral ratios, and varying algorithmic mechanisms) as highlighted in \cite{grobys2021stability, jeger2020analysis, zhao2021understand}. Furthermore, research should study the stability of stablecoins during crises not covered by current research, such as the 2018 Cryptocrash or the Celsius Network collapse. The COVID-19 crisis is studied only in \cite{jalan2021shiny} and using exclusively daily price data; utilizing higher frequency data can offer more insight into the sudden impacts of this crisis.

Moreover, none of the studies utilize blockchain transaction data. While price data is useful in measuring volatility, blockchain transaction data can offer insights into the behaviors that led to this volatility.

\subsection{Price Modeling}

\textit{\textbf{(11) What Are The Properties of Stablecoin Prices In Terms Of Linearity, Time-Variation, Regime-Switching, Normality, And Autoregressive Behavior, etc.?}}

Stablecoin price series exhibit autoregressive patterns, regime-switching behavior, non-linearities, heavy tails, and signs of spurious long memory.

Non-linearity and regime-switching are prominent characteristics of stablecoins' price behavior. \citet{maiti2020dissecting} found that, unlike many volatile cryptocurrencies, USDT's returns exhibit significant non-linearity, particularly during the COVID-19 pandemic. They also identified shifts between different market conditions, leading to the use of regime-switching models like Threshold Autoregressive (TAR) and Smooth Transition Autoregressive (STAR) to capture these dynamics. Similarly, \citet{pernice2021stablecoin} analyzed 11 fully collateralized stablecoins and observed a dual-regime behavior: one where price stability is maintained through arbitrage and another during extreme market conditions with more dramatic corrections. Their findings provide evidence of trend-reversion, where prices return to expected values through arbitrage mechanisms, modeled using the Caginalp-Balenovic asset flow dynamics model \cite{caginalp1999asset}.

In terms of long memory, \citet{ahmed2023long} discovered that Tether, USD Coin, and Binance USD exhibit spurious long memory. This suggests that apparent long-range autocorrelations in the data are driven by structural breaks rather than genuine persistence in the time series. 

Stablecoins also demonstrate heavy-tailed distributions, reflecting their susceptibility to extreme price movements. \citet{levene2021hypothesis} found that USDT, DAI, and USDC exhibit distributions with tails heavier than those of a Gaussian distribution, a common feature of financial time series influenced by extreme market conditions.

\insights{Stablecoin price time series demonstrate non-linearity, regime-switching behavior, autoregressive patterns, spurious long-memory, and heavy-tailed distributions.}

\textbf{Methodologies:} Researchers have employed a variety of methodologies to analyze stablecoin price behaviors. To detect non-linearity, \citet{maiti2020dissecting} and \citet{ahmed2023long} used the Brock-Dechert-Scheinkman (BDS) and Lagrange Multiplier tests. Stationarity has been examined using the Augmented-Dickey-Fuller Test (ADF) \cite{maiti2020dissecting}, though ADF's assumptions of linearity and absence of regime switches make it less suitable for stablecoins.

The presence of heavy-tailed distributions was analyzed by \citet{levene2021hypothesis} using Goodness-of-Fit tests, including the Jensen-Shannon Divergence and Kolmogorov-Smirnov Two-Sample tests. Long-memory properties were investigated in \cite{ahmed2023long} using a combination of Rescaled Range, Log Periodogram, Exact Local Whittle, and ARFIMA methods; this use of several methods is a good example of the required robustness. 

A range of methodologies were used to identify regime-switching and structural breaks. \citet{maiti2020dissecting} used Threshold and Smooth Transition Autoregressive models, while \citet{pernice2021stablecoin} employed fixed-effects panel regression and the Qu test. \citet{ahmed2023long} utilized Bai and Perron Structural Break Tests.

Despite the methodological rigor, three of the four studies (\cite{maiti2020dissecting, ahmed2023long, levene2021hypothesis}) relied solely on daily price data, which may obscure intraday volatility.

\textit{\textbf{(12) How Can Actors In Stablecoin Ecosystems Be Modeled Using Computational Simulations?}}
Only two papers address this research question, both authored by the same researchers \cite{bhat2021simulating, bhat2021daisim}. These studies introduce an open-source simulation tool that models investor buy/sell behavior within the MakerDAO stablecoin ecosystem. This tool can be used to study how DAI prices are affected by various parameters related to investors, MakerDAO, and collateral (ETH). 

\textbf{Methodologies: }The tool models investors as maintaining portfolios that include DAI, with decisions to buy, sell, mint, or burn DAI influenced by Markowitz's Optimal Portfolio Theory. These decisions affect the price of DAI through changes in supply and demand. Key parameters impacting the optimal portfolio include the expected returns of USD, ETH, and DAI; the covariance matrix of these assets; investors' risk tolerance; and MakerDAO's interest rates, liquidation ratio, and transaction fees.

While the model offers insights into the mechanisms that stabilize the DAI stablecoin, it has some limitations. Firstly, the model is designed to analyze situations where one parameter is varied while others remain constant; in reality, parameters can vary together, creating a combined effect. Secondly, the model assumes a single-collateral DAI, which is outdated as the platform has transitioned to a multi-collateral system. Thirdly, the demand and supply of DAI can be impacted by factors not included in the model, such as competition with other stablecoins.

\journalColorText{\textit{\textbf{(13) What Are the Most Effective Methods for Modeling and Forecasting Stablecoin Prices?}}}

\journalColorText{The main results of the following section are the following: the Time-varying Parameter Double Autoregressive (tvDAR) model effectively addresses USDT's time-varying volatility and persistence of volatility shocks; a positive Lyapunov exponent indicates that volatility shocks grow exponentially in USDT; the Scaled Muth-ARMA Model (sMuth-ARMA) model is more effective than Gaussian ARMA at capturing extreme events and volatility in USDT; the LSTM with Graph Convolution Network (LSTM-GCN) model outperforms simpler LSTM model in forecasting USDT prices, especially when the Financial Stress Index is included as an input; some Artificial Neural Network models didn't achieve high accuracy in forecasting stablecoin prices, but the sigmoid activation function yields the best results; simple rule-based trading algorithms demonstrate profitability on stablecoin prices.}

\journalColorText{\citet{djogbenou2023time} applied a time-varying parameter Double Autoregressive (tvDAR) model to Tether (USDT). This model is used to deal with USDT's time-varying volatility, persistence of volatility shocks, and autoregressive nature (for instance, periods of high volatility are more likely to be followed by high volatility). The authors also introduce a new stability measure for stablecoins, using the Lyapunov exponent. This exponent measures how prone a time series is to explosive instability. If this exponent is positive, then shocks to volatility grow with time, causing exponential volatility.}

\journalColorText{\citet{nascimento2023scaled} introduced a novel time series model called the scaled Muth-ARMA (sMuth-ARMA) for stablecoin price analysis. The model is an extension of the autoregressive moving average (ARMA) process but uses the scaled Muth law as its marginal distribution, improving the modeling of non-Gaussian data. Through Monte Carlo simulations, they show that the sMuth-ARMA model outperforms Gaussian ARMA when applied to Tether (USDT) stablecoin prices. Their model is better able to capture extreme events and volatility.}

\journalColorText{\citet{yin2024forecasting} integrated Long Short-Term Memory (LSTM) with Graph Convolution Network (GCN). In addition to USDT prices, they include the Financial Stress Index as an input to their model to account for the interconnectedness between cryptocurrency and traditional financial markets. Their findings demonstrate that their model outperforms univariate and multivariate LSTMs. In addition, they found that including the Financial Stress index significantly improves the model's accuracy.}

\journalColorText{\citet{Benitez2021} used Artificial Neural Networks (ANNs) to forecast stablecoin prices. Even though their ANN model did not produce highly accurate results, they found that the sigmoid activation function consistently produced the best results compared to linear and tanh functions. Note that their finding does not eliminate the usefulness of ANN models in forecasting stablecoin prices. For instance, \citet{yin2024forecasting} found that LSTM, a type of ANN, is useful.}

\journalColorText{\citet{baugci2024symmetric} designed stablecoin trading algorithms that buy and sell when the price deviates from the intended peg. The first algorithm buys and sells prices symmetrically around the expected price (e.g.: if the intended price is 1 then the buy price might be 0.99 and the sell price 1.01). The second algorithm allows asymmetric buying and selling prices (e.g., the buying price can be 0.99, while the selling price is 1.02). Their algorithms were tested on three stablecoins using high-frequency minute-by-minute data and found to be profitable.}


\journalColorText{\textbf{Methodologies: }Different methodologies have been used to model and forecast stablecoin prices: autoregressive models, neural networks, and rule-based algorithms.}

\journalColorText{Autoregressive models account for the fact that financial time series tend to exhibit autocorrelation (past prices are a predictor of future prices). \citet{djogbenou2023time} chose a time-varying Double Autoregressive (tvDAR) model. This specific model accounts for time-varying changes in volatility, which is realistic for time series whose volatility is impacted by varying market conditions. \citet{nascimento2023scaled} used a variant of the Autoregressive Moving Average (ARMA) model. The ARMA model takes into consideration two factors: the dependency of the current price on its previous values and the error of previous predictions. They specifically use the Scaled Muth–ARMA model (sMuth-ARMA). The sMuth-ARMA utilizes a Muth distribution instead of a normal distribution, which is more suitable for modeling heavy-tails. In general, financial time series tend to be more heavy-tailed, as extreme price movements are more likely than the probabilities given by a Gaussian distribution. The sMuth–ARMA model also introduces a variance control parameter, which allows the model to dynamically adjust the variance of the time series, depending on market conditions.}

\journalColorText{Neural networks can have an advantage over the previous econometric methods. They can be better for capturing non-linear relationships and handling complex data without requiring assumptions to be made. \citet{yin2024forecasting} combined a Long Short-Term Memory (LSTM) model with a Graph Convolutional Network (GCN). LSTM is a deep learning model useful for capturing complex dependencies between the current price and previous prices. GCN is a deep learning model utilized to capture interconnectedness between the price and different variables, represented in graph form.}

\journalColorText{\citet{baugci2024symmetric} forecasted stablecoin prices using a very simple rule-based model that assumes that stablecoin prices will revert to the peg. When stablecoin prices fall below the peg, the model predicts that it will go up, and vice versa. In general, this type of simple model is not useful for financial time series due to their lack of stationarity; however, stablecoins are supposed to be more stationary around the peg.}

\textbf{Research Gaps:} Authors studying stablecoin price modeling highlight several avenues for future research. \citet{maiti2020dissecting} emphasized the need to identify the most effective nonlinear models, particularly during crises when stablecoin price processes are subject to shocks. Additionally, there is limited use of machine learning methods in stablecoin modeling, with only two papers (\cite{yin2024forecasting, Benitez2021}) exploring this approach, leaving ample room for experimentation. \journalColorText{The two studies focus on ten stablecoins in total using ANN and GNN models; future research can experiment with more advanced architectures of the same models or with different models. The variables included are also limited to stablecoin prices and the financial stress index; future research can include more variables. Future research can also test the profitability of such models.}

Arbitrage opportunities in stablecoin markets also remain underexplored. \citet{baugci2024symmetric} demonstrated that a simple rule-based model was profitable, but this was the only study investigating arbitrage profits from trading stablecoins and did not include periods of market shocks. Future research could focus on developing more sophisticated trading algorithms and testing their performance under varying market conditions.


\section{Technical Analysis}
\label{sec:technical_analysis}
\journalColorText{While economic dynamics determine the market behavior and adoption of stablecoins, their functionality depends on a technical architecture involving blockchain technology, cybersecurity, and system-level design choices. For instance, stablecoins rely on smart contracts that encode issuance, redemption, consensus protocols, and stabilization mechanisms. Stablecoin systems also often need to be technologically integrated with the broader DeFi infrastructure, such as decentralized exchanges and price oracles. This section synthesizes academic research within the technical analysis macro-category, encompassing four sub-categories: system design and implementation, stabilization mechanisms, failure analysis, and security and privacy considerations.}

\subsection{Design And Implementation}


\textit{\textbf{(14) How Can We Improve The Design And Implementation of Stablecoins?}}

\begin{figure}[htbp] 
    \centering
    \includegraphics[width=0.48\textwidth]{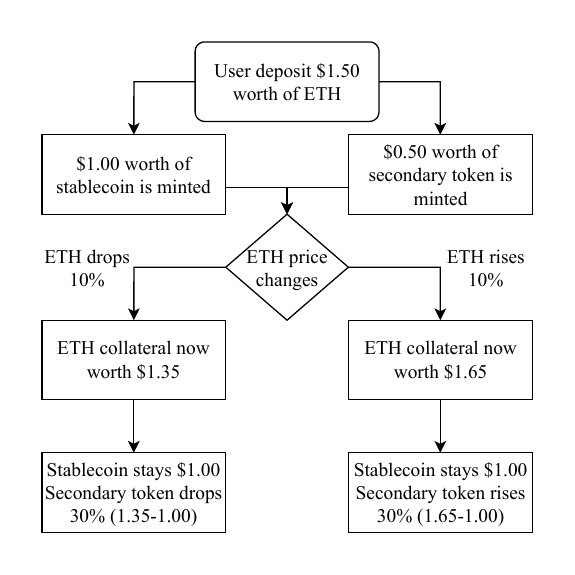}
    \caption{\journalColorText{Structure of a dual-token stablecoin mechanism. Dual-token stablecoin mechanism based on overcollateralized ETH. A user deposits \$1.50 in ETH to mint \$1.00 of stablecoin and \$0.50 of a secondary token. If ETH changes ±10\% in value, the stablecoin maintains its peg while the secondary token absorbs the volatility, with leveraged exposure to collateral fluctuations. Stablecoin holders will be asked to liquidate if ETH drops by more than 33\% because the value of the \$1.50 collateral would fall below \$1.00, making it insufficient to fully back the stablecoin.}}
    \label{fig:dual-token}
\end{figure}

The main proposals for improving the design and implementation of stablecoins include the following: pegging a stablecoin to a basket of fiat currencies; using physical diamonds as collateral; adjusting the exchange rate to deter speculative attacks; developing novel stablecoins that combine a high-risk token with a low-risk stablecoin; modifying minting costs to influence supply; issuing bonds to regulate supply; designing a stablecoin tailored for radio spectrum trading; and relying on arbitrageurs and third-party investors to mitigate collateral risk.

\citet{giudici2022libra} proposed a stablecoin design that is not pegged to a single currency but to a basket of fiat currencies and demonstrates the increased stability of this approach. \citet{asadov2023toward} analyzed the history of several precious metals and found that a stablecoin backed by a combination of gold-silver (88:12 ratio) optimizes stability. \citet{bandara2022gemcash} introduced a system backed by redeemable physical diamonds, while \citet{tsai2022new} suggested a stablecoin design where the collateral is held by custodian banks and certificates of ownership are traded on the blockchain.

\citet{dong2019elasticoin} introduced Elasticoin, a stablecoin system that uses Proofs of Sequential Work (PoSW) to reduce volatility by dynamically adjusting the minting cost. The system adjusts the computational effort required to mint in order to increase or decrease the supply of minted coins.

\citet{routledge2022currency} proposed a stablecoin design that prevents speculative attacks. A speculative attack is when traders heavily sell a currency because they expect its value to fall, leading to a self-fulfilling prophecy. Their solution employs a dynamic exchange rate adjustment, temporarily lowering the peg to make attacks more costly. Using formal mathematical proofs, the authors demonstrate that this policy reduces the likelihood of speculative attacks. Unlike conventional approaches, which aim to push prices toward the peg, this method strategically moves the exchange rate away from it to stabilize the system.

\citet{heinonen2021creation} proposed a stablecoin system with two wallets: Investment and Saving wallets. The Investment wallet absorbs more risk. If demand increases and the price of the stablecoin rises above the peg, the system increases the coin supply, distributing it proportionally to users’ Investment wallet balances. Conversely, if demand decreases, the system reduces supply by deducting from the Investment wallet balances. To address cases where users lack sufficient funds to absorb reductions, the authors introduce ''antimoney”, a credit mechanism allowing negative balance, capped by the user’s Saving wallet balance.

\citet{kazemian2022frax} introduced FRAX, a stablecoin that transitions from full to partial collateralization as adoption grows. A secondary token, FXS, absorbs volatility during the partial collateralization phase. Users mint FRAX by depositing less than \$1 in collateral and supplementing it with FXS. The percentage of collateral vs. FXS needed depends on a dynamic ratio. This ratio adjusts to maintain stability: if FRAX trades above the peg, the ratio is lowered, allowing more FRAX to be minted and vice versa. Moreover, as demand for FRAX grows, the need for FXS increases, raising its price and rewarding stablecoin holders.

\citet{zahnentferner2023djed} also proposed a system where a secondary token represents the collateral value in excess of \$1. However, the system relies on excess reserves to mitigate against potential collateral price drops. 
Moreover, the authors propose that the system temporarily restricts stablecoin sales in extreme market scenarios, allowing time for collateral prices to recover.

\citet{das2023incentivized} proposed a fiat-pegged stablecoin system collateralized with Ethereum (ETH). However, unlike the previous proposal in \cite{zahnentferner2023djed}, collateral in excess of \$1 is not provided by the stablecoin owners but by external investors. These investors assume the collateral's volatility risk while benefiting from its potential appreciation and also earning a share of transaction fees. Upon redemption,  collateral value in excess of \$1 is returned to these investors. This concept closely resembles the red-black coin system proposed by \citet{salehi2021red}.

\citet{zhang2023enhancing} introduced SCoin, a stablecoin system for blockchain-based trading in the radio frequency spectrum. The Spectrum Resource Management Center (SRMC) issues SCoins and manages spectrum transactions through smart contracts. Spectrum Trading Service Providers (STSPs) obtain spectrum rights from SRMC to trade with users. Access nodes connect users via various infrastructures like satellites and mobile networks. A Delegated Proof of Stake mechanism validates transactions, considering both SCoin holdings and spectrum interference, discouraging abuse of the spectrum by large coin holders.

After analyzing the previous stablecoin proposals, we noticed that some designs share significant similarities. For instance, both \citet{tsai2022new} and \citet{bandara2022gemcash} proposed asset-backed stablecoins using redeemable physical collateral stored in custodian banks, differing primarily in the type of asset (fiat currency versus diamonds). 

Moreover, several authors proposed a stablecoin that utilizes a secondary volatile token that represents residual collateral value \cite{kazemian2022frax, zahnentferner2023djed, das2023incentivized}. However, they have some differences. For example, FRAX \cite{kazemian2022frax} is the only one that transitions to partial collateralization. And in \cite{das2023incentivized}, external investors provide the excess collateral. Both the proposals in \cite{kazemian2022frax} and \cite{zahnentferner2023djed} have been implemented and received notable market success. The structure of this dual-token mechanism has been illustrated in Fig.~\ref{fig:dual-token}.

Some of the proposals are also similar to stablecoins that have been previously implemented. For instance, the basket-backed stablecoin proposed in \cite{giudici2022libra} is similar to the idea behind Libra (renamed Diem). Similarly, the dual-token mechanisms in \cite{zahnentferner2023djed, heinonen2021creation, kazemian2022frax} share features with MakerDAO's DAI, though they introduce some uniqueness. Additionally, some papers (e.g., \cite{liu2020mover}) propose well-established mechanisms like arbitrage, open-market operations, and bond issuance.

\textbf{Methodologies:} 
This section employs different methodologies, including economic modeling, blockchain implementation, formal verification, and scenario analysis.

Stablecoins pegged to a basket of currencies rely on the comovements of the basket's currencies for stability. Techniques like variable decomposition and vector autoregression (VAR) help analyze these comovements by assessing how changes in one currency affect others \cite{giudici2022libra}. A basket with highly correlated currencies provides less diversification, reducing stability per Modern Portfolio Theory \cite{markowitz1952modern}. Researchers can optimize risk by factoring in each currency's historical variance and their covariances. Quantifying risk can be done using measures such as variance or Value-at-Risk \cite{liu2020mover}. 

In the design of stablecoins, game theory is a useful tool, modeling the behavior of relevant actors and incentivizing them to behave in ways that serve stablecoin stability. For example, game theory has been used to analyze speculative attacks with a system of traders and attackers \cite{routledge2022currency}. Attackers can overwhelm the supply of stablecoins in a short period of time, destabilizing it. However, they show using game-theoretic modeling that their design would make it unprofitable for attackers to do so. While game-theoretic models provide a logical analysis framework, they are often built on assumptions that oversimplify real-world scenarios. For instance, the model built in \cite{routledge2022currency} assumed that traders' behaviors are independent, ignoring herding behaviors. 
The model also ignores certain practical phenomena, including deleveraging spirals, which might occur due to the proposed de-pegging mechanisms.

\insights{Game theory models related to stablecoins assume independent traders, ignoring herd behaviors. They also overlook other relevant phenomena such as the effects of deleveraging spirals or arbitrageurs.}

The proposed implementation of stablecoins takes blockchain and smart contract design into consideration. Some researchers implemented innovative algorithmic stability rules using Ethereum smart contracts \cite{routledge2022currency, das2023incentivized}. Ethereum is the most popular platform, providing many heavily-tested tools, an active developer community, and wide integration with DeFi protocols. \citet{zahnentferner2023djed} used Cardano and Ergo to implement their stablecoin design. These platforms help formally verify the reliability of their implementations. \citet{dong2019elasticoin} utilized Ergo's blockchain, which is suitable for implementing innovative consensus mechanisms. \citet{bandara2022gemcash} utilized a permissioned blockchain, allowing more centralized control.

To test the stability of the proposed stablecoins against different possible scenarios, synthetic data has been generated using techniques like Monte Carlo simulations \cite{liu2020mover}. Some researchers also utilized formal verification or mathematical proofs to illustrate the level of stability. For example, \citet{zahnentferner2023djed} used bounded model checking and interactive theorem proving with Isabelle (a higher-order logic theorem proof assistant) to demonstrate the stability of their design.

\journalColorText{\textit{\textbf{(15) How are CBDCs designed, and how should they be designed?}}}

\journalColorText{The main results of this section are the following: CBDCs should coexist with traditional cash; central banks prefer permissioned/private blockchains for implementing CBDCs; CBDCs can be programmed to prevent irresponsible printing of money, mitigating risk of hyperinflation; a cash-like CBDC model is recommended for Indonesia, maintaining anonymity of users; a centralized blockchain system with commercial banks as intermediaries is recommended for CBDCs, particularly in Mexico; challenges facing Cambodia's Bakong CBDC include digital payment adoption, risk of disintermediation of banks, and maintaining the required USD reserves.}

\journalColorText{\citet{morales2021implementing} analyzed three pilot CBDC projects, particularly the Sand Dollar (Bahamas), the e-Peso (Uruguay), and the e-Krona (Sweden). Through their analysis, they argue that CBDC systems should be designed to coexist with cash and involve partnerships with private payment service providers. This can help foster innovation while maintaining central bank control. They also argue that security and 24/7 availability are essential for CBDCs. In addition, CBDCs should maintain interoperability with current payment systems, domestically and internationally, such as bank transfer systems, point-of-sale systems, and digital payment methods. Ideally, central banks would collaborate to support cross-border transactions using CBDCs, enabling efficient international payments. CBDCs should be made accessible to users with low technological knowledge. CBDCs should be interoperable with domestic and international compliance systems that manage anti-money laundering (AML), know-your-customer (KYC) protocols, and other financial regulations. The paper's authors do not go into technical details about how their recommendations should be implemented.}

\journalColorText{\citet{sethaput2023blockchain} explored how distributed ledger technology (DLT) can be used to facilitate CBDCs, discussing both public and permissioned blockchains. They provide an analysis of various CBDC projects around the world. The authors discuss the characteristics of 17 CBDC projects from different countries. For instance, the authors note that central banks prefer permissioned or private DLT platforms to implement CBDCs, as they offer stricter privacy, scalability, and more control.}

\insights{\journalColorText{The interoperability of a CBDC with existing financial systems (e.g., bank transfers, POS systems, and international payment networks) is crucial for widespread adoption.}}

\journalColorText{\citet{zams2020designing} explored the design of a CBDC for Indonesia, concluding that a cash-like CBDC model is the most appropriate for Indonesia. Their proposed model resembles traditional cash because it is universal, anonymous, peer-to-peer, and non-yield-bearing. Authors argue that their model will promote financial inclusion and reduce shadow banking while maintaining efficiency and security.}

\journalColorText{\citet{marthinsen2022hyperinflation} proposed creating a hyperinflation alternative cryptocurrency (HACC) to combat hyperinflation faced by countries. This is based on an analysis of historical hyperinflation cases in several countries. The key objective is to put controls on central banks that might execute policies exacerbating hyperinflation (mainly excess issuance of currency). This is primarily achieved through a design that operates independently of central banks on a decentralized ledger. \citet{marthinsen2022hyperinflation} discussed 16 points related to the design of HACCs. Ideally, HACCs would have 100\% reserve backing or a fractional reserve system with guarantees from trusted international organizations. Consensus protocols, such as Proof-of-Stake (PoS), could be used to ensure security, decentralization, and scalability. Scalability and low transaction fees are especially important for CBDCs due to the vast amount of financial transactions performed within a country. Privacy should not precede anonymity, as governments need to hold users accountable for illegal transactions. The source code of HACC should be open and transparent to earn the highly needed trust of governments facing hyperinflation. HACCs should be accessible and interoperable with traditional payment systems and global financial systems.}

\insights{\journalColorText{Decentralized CBDC models are especially useful for economies with distrust in central banking systems.}}

\insights{\journalColorText{A key advantage of programmable CBDCs is their ability to restrict excessive money issuance, mitigating risks of hyperinflation.}}

\journalColorText{\citet{flores2023cbdc} analyzed the development of a CBDC for Mexico. They examined several Mexican peso-pegged stablecoins, noting that MMXN (Moneta Digital) is the most widely adopted. To create a successful CBDC, several challenges must be addressed, including the role of commercial banks, the risk of disintermediation, and the need to improve financial literacy. Similar to other studies, they recommend that the CBDC coexist with traditional cash. The authors present three implementation proposals, identifying a centralized design as the most feasible for initial implementation.}

\insights{\journalColorText{A hybrid system, where CBDCs coexist with physical cash, is recommended.}}

\journalColorText{The first proposal involves a centralized model issued on a blockchain, with commercial banks serving as intermediaries. The second proposal also suggests a centralized system, but instead of a token-based model, it employs an account-based structure. In this model, CBDC holders maintain accounts either directly with the Central Bank of Mexico or with trusted intermediaries. While this approach aligns more closely with the current banking system, it offers less innovation potential. The third proposal is a decentralized token-based system, which increases transparency but presents disadvantages in terms of central bank control.}

\journalColorText{\citet{ueda2024design} studied the implementation of Cambodia's CBDC, the Bakong, launched in 2020. Bakong is designed to operate in Cambodian Riel and also USD; this is relevant to Cambodia's financial system, which has around 80\% of bank deposits in USD. The Bakong uses the Hyperledger Iroha blockchain platform. Retail users can only access this CBDC through commercial banks. The authors note some challenges that need to be addressed, including limited digital payments in Cambodia, disintermediation risks (where people prefer Bakong over bank deposits), and Cambodia's ability to back the CBDC (particularly the USD Bakong). They note that the Bakong design is intended for wholesale interbank transfers. The authors believe that the risks of this CBDC outweigh the potential benefits.}

\journalColorText{\citet{genc2024literature} performed a literature review on the design and implementation of CBDCs. They note increasing interest in CBDCs and China's CBDC development. The review introduces various design models, including retail versus wholesale CBDCs, direct versus indirect, account-based versus token-based systems, and their associated benefits and risks. Retail CBDCs are designed for use by the public, while wholesale CBDCs are restricted to financial institutions. A direct model involves individuals holding accounts directly with the central bank, while an indirect model works through intermediaries like commercial banks. Account-based CBDCs are tied to the holder's identity, while token-based CBDCs allow for transfers without linking to personal identities (similar to cash). The authors argue that wholesale CBDCs have the advantage of reducing bank disintermediation. They also argue that account-based models using centralized ledgers offer better security and government surveillance. They also stress the importance of interoperability with domestic and international payment systems.}

\insights{\journalColorText{Central banks generally favor permissioned blockchains due to higher control.}}

\journalColorText{Some conflicting approaches are notable among different CBDC designs. For instance, \citet{marthinsen2022hyperinflation} recommended a decentralized approach with limited central bank control to gain investors' trust, especially in countries with distrusted central banks or facing hyperinflation. In contrast, the authors of \cite{flores2023cbdc, sethaput2023blockchain, genc2024literature} advocated for a more centralized design to enhance security and legal oversight. This demonstrates a trade-off between central bank oversight over the CBDC and investors' trust that the central bank will not mismanage it. In certain circumstances—such as economies with hyperinflation or distrust in central banking institutions—gaining users' trust may take precedence over maintaining central bank control.}

\journalColorText{Another trade-off exists between anonymity and legal accountability. For example, the authors of \cite{ueda2024design, zams2020designing} recommended a cash-like CBDC model that emphasizes user anonymity. On the other hand, the authors of \cite{flores2023cbdc, sethaput2023blockchain, genc2024literature} proposed options such as permissioned blockchains or bank-account-based CBDCs, prioritizing government control and legal accountability. Some proposals aim to find a balance, for instance, allowing users to make anonymous transactions using vouchers only for small amounts \cite{tronnier2021privacy}.}

\journalColorText{\textbf{Methodologies: }For research related to the design and implementation of CBDCs, different methodologies have been used. Some researchers interviewed stakeholders in CBDC pilot programs to understand the motives, design choices, regulatory adaptations, operational aspects, etc. \cite{morales2021implementing}. Others interviewed industry experts who are not directly related to any CBDC pilot program but nevertheless have valuable insights regarding CBDCs, particularly for a specific country (e.g., Indonesia \cite{zams2020designing}). Interviews can follow systematic methodologies such as the Delphi Method or the Analytic Network Process \cite{zams2020designing}. Other researchers surveyed users instead of experts. Surveying users can provide insights into their preferences related to CBDCs \cite{ueda2024design}. Hypotheses such as the willingness of users to adopt CBDCs can be tested by surveying users and applying statistical tests to the survey results \cite{ueda2024design}. Instead of interviewing, some authors relied purely on literature review to back their arguments \cite{flores2023cbdc}.}



\journalColorText{\textit{\textbf{(16) How can certain stablecoin-related services be designed and implemented?}}}

\journalColorText{The main proposals of this section are the following: a blockchain-based stablecoin system where the issuer holds ETFs as a custodian; a more efficient sharding method for unspent transaction outputs (UTXO) in token-based systems; a centralized bridge infrastructure to enable transfers between different blockchains; a mechanism where stablecoin holders are prevented from selling during periods of extreme market stress.}

\journalColorText{\citet{ciriello2021tokenized} proposed a system called Tokenized Index Fund (TIF). These systems are basically blockchain-based stablecoins, where the issuer of the stablecoin holds the ETFs as a custodian. This system provides several benefits, including more efficient transfers of ETF ownership and fractionalization of ETF shares.}

\journalColorText{\citet{jin2022utxo} proposed a more efficient Unspent Transaction Output (UTXO) sharding method aimed at improving the scalability of token-based systems, especially stablecoins. Sharding is the process of dividing a blockchain network into smaller units called shards. Each shard processes its transactions independently, which helps increase throughput since multiple transactions can be processed in parallel across different shards. A challenge that increases latency, however, is cross-shard transacting, where transactions are sent between users assigned to different shards. The proposed method aims to minimize the occurrence of cross-shard transactions by placing accounts likely to transact together frequently in the same shard. The authors also propose transaction batching and asynchronous cross-shard communication mechanisms to reduce the impact of cross-shard transactions.}

\journalColorText{\citet{machado2024architecture} designed and implemented a stablecoin service that enables cross-chain interoperability. The system uses a centralized bridge solution, which involves burning tokens in the originating chain and minting equivalent tokens on the target chain using smart contracts.}

\journalColorText{\citet{dudekula2023pioneering} detailed a design of a simple application used for international bank transfers. The fiat currency to be sent is first converted into stablecoins, then transferred to a wallet in the receiving country on-chain, before being converted into fiat currency again to reach the receiver.}

\journalColorText{\cite{islam2022resilience} proposes a method to improve the resilience of a peer-to-peer (P2P) energy trading platform. The platform operates by receiving fiat currency from consumers in exchange for stablecoins. Consumers can then use these stablecoins to purchase products from producers on the trading platform. Producers are allowed to hold the stablecoins for a defined period of time, denoted as \(M_t\). In scenarios where more consumers join the platform, the platform must ensure it has a sufficient supply of stablecoins to distribute. If a shortage of stablecoins occurs, the platform can reduce the holding period \(M_t\), thereby requiring producers to return their stablecoins in exchange for fiat currency. This holding period is dynamically adjusted to maintain a Liquidity Coverage Ratio (LCR) close to 1. The LCR is calculated as the ratio of the current stablecoins in reserve to the expected demand for stablecoins by consumers over the next 30 days. This metric is inspired by the Basel III accord, which aims to enhance the resilience of traditional banks.}

\journalColorText{\textbf{Methodologies: }All five papers propose a conceptual framework for their proposed stablecoin-related system. \citet{ciriello2021tokenized} conceptually combined tokenized securities and index funds to propose a framework for Tokenized Index Funds (TIFs). \citet{dudekula2023pioneering} proposed a framework for using stablecoins for remittances and presents a simple flow of assets on- and off-chain to demonstrate this process.}

\journalColorText{Some papers implement their proposal. \citet{jin2022utxo} implemented their proposed sharding method on Corda, a permissioned blockchain platform. \citet{jin2022utxo} included experimental results quantifying the linear scalability and low latency of their implemented system. \citet{islam2022resilience} implemented their proposed mechanism with the open-source permissioned blockchain platform Hyperledger Fabric. Using smart contracts, the authors automate adjustments to the stablecoin holding times in response to reserve and expected outflow fluctuations. They run simulations comparing their mechanism with other mechanisms, analyzing how effectively the mechanisms sustain reserve stability under stress.}

\journalColorText{\citet{machado2024architecture} presented a microservice-based architecture for a stablecoin backend system that involves cross-chain interoperability. The microservice approach breaks down the stablecoin service into smaller, independent units. Key microservices include an API Gateway, a webhook that listens to events from the banking service, a smart contract interface to interact with the blockchain, a component responsible for storing permanent data like user details, a component that facilitates communication between microservices, a microservice notifying users about their account activity, a microservice displaying the total reserves, and a microservice that caches frequently accessed data like gas prices. The proposed stablecoin uses the ERC-20 standard. A centralized bridge is employed for cross-chain functionality. This allows users to move stablecoins between different blockchains (Ethereum and Polygon in the proof-of-concept) without needing to convert back to fiat currency. The bridge operates by burning tokens on the source chain and minting an equivalent amount on the destination chain.}

\textbf{Research Gaps:} There is significant potential for novel stablecoin proposals, including those based on flexible pegs (e.g., RAI \citep{gebfoundation2024reflexer}) or baskets of assets, as highlighted in \cite{giudici2022libra}. Privacy and security remain underexplored areas, particularly for stablecoins and CBDCs \cite{giudici2022libra, morales2021implementing, sethaput2023blockchain}. Assessing the security levels of different stablecoin designs is crucial, as is the development of rigorous formal verification methods and standardized security audits for stablecoins to reduce the risk of exploits. Furthermore, a framework is needed to understand the tradeoffs between security, privacy, transparency, efficiency, and regulatory oversight. In this regard, reconciling user privacy with AML/KYC compliance is an essential challenge. Proposals for privacy-preserving stablecoins could address this gap.

Additionally, more robust scenario analysis and stress testing are necessary to uncover potential failure points before they arise in practice. Developing a standardized testing framework would facilitate this. Stability assessments should also consider varying levels of adoption, as some authors, such as \citet{kazemian2022frax}, argue that greater adoption enhances stability.

The lack of regulatory and industry standards compliance is a significant challenge for stablecoins, as highlighted in \cite{eichengreen2022stablecoins}. Developing proposals that align with these standards could enable broader adoption, particularly among institutions that require strict compliance. 

Another future research avenue is related to CBDCs. CBDCs carry the promise of revolutionizing monetary policy through novel monetary policy tools such as programmable negative interest rates, expiration dates on money, and smart contract-based taxation. Research on the design and implementation of these tools is needed.

Moreover, most authors proposing stablecoin designs focus on full or partial collateralization. Fully algorithmic stablecoins, however, remain underexplored despite offering key advantages such as reduced reliance on trusted custodians and greater capital efficiency.

\subsection{Stablecoin Mechanisms}

\textit{\textbf{(17) What Are The Characteristics Of Stabilization Mechanisms Used In Stablecoins?}}

\begin{table*}[htbp]
  \caption{\journalColorText{Classification of stablecoin designs based on stabilization mechanisms.}}
  \label{table:stablecoin_classification}
  \centering
\begin{tabular}{@{} l l >{\raggedright\arraybackslash}p{6cm} l @{}}
    \toprule
    \textbf{\journalColorText{Category}} & \textbf{\journalColorText{Sub‐category}} & \textbf{\journalColorText{Main Mechanism(s)}} & \textbf{\journalColorText{Example(s)}} \\
    \midrule
    \multirow{2}{*}{\journalColorText{Collateralized}}
      & \journalColorText{Off‐Chain}
        & \journalColorText{Custodial 1:1 reserves; on‐demand redemption.}
        & \journalColorText{Tether (USDT), USD Coin (USDC)} \\
      & \journalColorText{On‐Chain}
        & \journalColorText{Over-collateralized with automated liquidations.}
        & \journalColorText{Dai (DAI), Liquity USD (LUSD)} \\
    \midrule
    \journalColorText{Algorithmic}
      & \journalColorText{—}
        & \journalColorText{Elastic‐supply rebases and with dual-token seigniorage.}
        & \journalColorText{Ampleforth (AMPL), Empty Set Dollar (ESD)} \\
    \midrule
    \journalColorText{Hybrid}
      & \journalColorText{—}
        & \journalColorText{Fractional collateral ($<100\%$) plus algorithmic mechanisms.}
        & \journalColorText{Frax (FRAX)} \\
    \bottomrule
  \end{tabular}
\end{table*}

Stablecoins are either fully collateralized, partially collateralized, or non-collateralized (algorithmic), \journalColorText{as classified in Table~\ref{table:stablecoin_classification}}. Stablecoins that are not fully collateralized use mechanisms like supply adjustments, dynamic fees, two-token mechanisms (e.g., seignorage shares), dynamic mining rewards, bond issuance, dynamic collateral ratios, prevention of selling, or automated buying/selling to impact the supply and demand of the stablecoin.

Fifteen papers analyze the stabilization mechanisms used in stablecoins. The simplest mechanism is full collateralization. Full collateralization means that users must deposit assets equal to or greater than the peg value with a custodian in order to receive stablecoins. 
For example, a user might deposit \$1 USD or its equivalent in Bitcoin to receive \$1 of a stablecoin. 
Using cryptocurrency as collateral for fiat-pegged stablecoins introduces volatility risk. This risk can be mitigated by requiring over-collateralization. If the value of the cryptocurrency collateral falls below a certain threshold, the system liquidates the collateral to maintain the peg. The most popular stablecoin utilizing such a mechanism is DAI \cite{moin2020sok, kjaer2021empirical}.

Partially collateralized stablecoins reduce the reserve ratio, holding, for example, \$0.70 for every \$1 issued. This lower collateralization allows issuers to invest the remaining funds in more volatile assets, offering potential rewards but introducing additional risk. Under-collateralized and non-collateralized stablecoin systems rely on algorithmic mechanisms to ensure stability.

Algorithmic stability mechanisms noted in the reviewed papers include seigniorage shares, increasing supply by issuing coins, supply rebase, dynamic transaction fees, dynamic lending rates, dynamic mining rewards, automated market operations, issuing bonds, dynamic collateral ratios, and preventing users from selling for a limited period of time \cite{mita2019stablecoin, pernice2019monetary, gadzinski2023stablecoins, koutsoupakis2020stable}.

One prominent algorithmic mechanism is the use of seigniorage shares. In this context, seigniorage refers to issuing secondary tokens related to the stablecoin to absorb volatility \cite{moin2020sok, salehi2021red}. For example, in the red-black model \cite{salehi2021red}, users deposit volatile cryptocurrencies and receive stablecoins and ``red'' tokens. The red tokens represent the excess value of the collateral. If the price of the collateral increases, red token holders benefit from amplified returns. The red tokens represent the excess value of the collateral. To illustrate, assume the ETH collateral deposited is worth \$1.50; the user gets \$1.00 worth of stablecoins and \$0.50 worth of red tokens. If the price of ETH increases by 10\%, then the collateral deposited would be worth \$1.65. The value of the stablecoin remains \$1.00, while the red coin becomes worth \$0.65 (an increase of 30\%). If the collateral price decreases significantly, the owners of red coins can lose all their investment, while the stablecoin owner will be forced to liquidate the collateral, losing nothing in the process. Secondary tokens can grant their holders a share in the profit of the stablecoin system and/or voting power. 

Another algorithmic mechanism is supply rebase, which refers to a mechanism where stablecoins are burnt automatically to decrease supply and increase the price \cite{pernice2019monetary, moin2020sok}. Stablecoins can be burnt from all or a subset of users, depending on the specific stablecoin system.

Fees can also be automatically adjusted to impact supply and demand. For example, increasing transaction fees can discourage transfers, reducing demand. Other fee types include minting fees, withdrawal fees, and borrowing-related fees, all of which can influence supply and demand \cite{mita2019stablecoin}. For instance, decreasing the fees associated with borrowing (e.g., loan-origination fees) can increase demand for the stablecoin and raise prices. At the protocol layer, systems can dynamically adjust mining costs. Increasing mining costs reduces the incentive to mine new stablecoins, thereby limiting supply \cite{mita2019stablecoin}.

Automated market operations (AMO) refers to the buying or selling of assets on the market to influence the price of the stablecoin \cite{mita2019stablecoin, pernice2019monetary}. For instance, the issuer can use its reserves to buy the stablecoin, pushing up its price. This can be done automatically with smart contracts.

Another used mechanism to stabilize algorithmic stablecoins is the issuance of bonds. If the price falls below the peg, the issuer can reduce the supply of stablecoins by issuing bonds that users can purchase using stablecoins \cite{moin2020sok}. These bonds promise future payment after the crisis subsides. This is analogous to how traditional central banks issue bonds to reduce money supply.

For partially collateralized stablecoins, the system can algorithmically change the required collateral to impact stablecoin supply \cite{moin2020sok}. Decreasing the required collateral would increase supply, as more stablecoins can be minted for the same amount of collateral provided by users.

These algorithmic mechanisms, however, are not without their risks. Studies highlight that algorithmic stablecoins often exhibit higher volatility compared to collateralized designs \cite{jarno2021does}. \citet{charoenwong2023computer} demonstrated theoretically that guaranteeing the stability of purely algorithmic stablecoins is not possible; this finding seems to go against other papers that show confidence in algorithmic stabilization mechanisms, including \cite{dong2019elasticoin}. Despite these challenges, algorithmic stablecoins are not without value, as a currency can still be useful even without absolute and guaranteed stability. Moreover, the majority of stablecoin proposals are not fully algorithmic. In this regard, \citet{heinonen2021creation} claimed to formally verify the stability of a partially collateralized system that integrates algorithmic mechanisms.


\insights{Stablecoins that are not fully collateralized use algorithmic mechanisms to control supply and demand.}

\insights{Despite their higher empirical instability and the theoretical impossibility of guaranteed stability, algorithmic stablecoins offer value through greater capital efficiency and no need for a trusted centralized custodian.}

\textbf{Methodologies: }The most common methodology used to examine stablecoins' stabilization mechanisms is to analyze stablecoin-related material (white papers and official documentation, project websites, blogs, etc.) \cite{pernice2019monetary, moin2020sok, clark2019sok, li2024stablecoin}. For example, \citet{pernice2019monetary} ``surveyed white papers, websites, and, when available, price data of 24 stablecoin projects'', classifying each stablecoin based on its type of collateralization, its exchange rate regime, and its use of various stabilization techniques. \citet{clark2019sok} analyzed 185 articles from CoinDesk, finding information about 25 stablecoins. Based on this information, they classified each stablecoin into one of five stability mechanisms.

Other papers examine stabilization mechanisms through empirical analysis. For instance, \citet{lyons2023keeps} examined arbitrage activity using blockchain transactions where users mint USDT from Tether treasury and sell it for profit in secondary markets, bringing the price of USDT down to the mint cost. Similarly, \citet{kjaer2021empirical} empirically studied the stabilization mechanisms of the stablecoin system MakerDAO. MakerDAO issues the stablecoin DAI and accepts ETH as collateral. If the price of ETH drops below the collateralization ratio, a liquidation function is automatically called. This function offers the ETH collateral for auction to repay the DAI to the system. These stabilization mechanisms are enforced through contract calls on the Ethereum blockchain. \citet{kjaer2021empirical} analyzed these contract calls and the movement of funds between contracts to assess collateralization ratios, liquidation events, and auction effectiveness.

\journalColorText{\textit{\textbf{(18) What Are The Characteristics Of Governance Mechanisms In Stablecoins, And How Do They Impact The Platforms?}}}

\journalColorText{The main results of this section are the following: different voting tasks (strategic vs. operational decisions) impact DeFi platforms differently; votes with knowledgeable participants have a more positive impact; more participation of large token holders in voting has negative impacts on the platform; bribery mechanisms are the most efficient way to impact voting outcomes, followed by participating in vote-pooling platforms, followed by direct holding of tokens; audits of stablecoins face challenges such as inconsistencies among different reports, conflict-of-interest issues, and short coverage periods; audits have a positive impact on DeFi platform value.}

\journalColorText{Many stablecoins rely on governance mechanisms where token holders vote on key decisions. These votes impact critical parameters such as changes in the stabilization algorithm, the amount of collateral required, how the collateral is invested, the different transaction fees and interest rates, upgrades to the protocol code, etc. \citet{zhao2022task} categorized MakerDAO voting tasks into strategic and operational tasks. Strategic tasks involve voting on decisions that are high-impact and long-term, such as onboarding new collateral types, updating software, or making personnel changes. They find that this type of voting task has a positive effect on price stability. Operational tasks, on the other hand, are made on short-term decisions such as modifying interest rates or transaction fees. Such decisions are found to precede more instability in the price. The reason for this could be that voting on such decisions is slow, resulting in the execution of outdated policies. Expectedly, the authors find that having voters with more knowledge improves the impact of voting. Less expectedly, however, authors find that votes with more large token holders' participation tend to have a negative impact. This leads the authors to recommend reducing the power of large token holders in the voting system and encouraging small voters to participate. Several papers shed light on the concentrated voting power a few large token holders have on the MakerDAO ecosystem, creating an imbalance and allowing these few holders to vote in their own interest rather than for the interest of the overall ecosystem \cite{sun2024decentralization, zhao2022task, kozhan2022fundamentals}.}

\insights{\journalColorText{The two stablecoin governance mechanisms discussed in the literature are voting and auditing.}}

\insights{\journalColorText{Concentration of voting power is a problem in some stablecoin platforms.}}

\journalColorText{\citet{lloyd2023emergent} analyzed a voting model used on the stablecoin-trading platform Curve Finance. The veToken (Vote-Escrowed Token) model requires stakeholders to lock tokens for a fixed period in exchange for voting weight. This approach gives more weight to voters who have a longer-term interest in the platform. However, several practices have emerged that undermine this model, particularly bribing voters through a platform called Votium. \citet{lloyd2023emergent} showed a very high correlation between the value of bribes and voting outcomes using the veToken model. They compare different methods of influencing voting outcomes and find that bribing is the most cost-efficient. This bribing model is made possible because voters can delegate their voting rights in exchange for payment. Moreover, votes are transparently recorded on the blockchain, allowing Votium to hold voters accountable and bribe them according to their votes.}

\insights{\journalColorText{Bribing users is a highly effective and cost-efficient method of influencing voting outcomes on stablecoin platforms.}}

\journalColorText{Financial auditing of stablecoins is done manually by independent parties who verify the stablecoin issuer's reserve maintenance and quality. This can involve verification of reserves by contacting banks or custodians holding them or reviewing the public ledger for reserves held on-chain \cite{moura2022ledger}. Auditors may also review transactions involving the issuance and redemption of stablecoins to confirm that the system is well-managed. In addition to this, auditors can verify that stablecoin issuers are following relevant regulations such as anti-money laundering policies and financial reporting standards.}

\journalColorText{Several challenges facing the auditing of stablecoins are highlighted by researchers. There is a lack of a unified standard approach to accounting and auditing stablecoins. This creates inconsistencies among audited reports of Tether, as noted by \cite{fernandez2024asset}. Auditors are also sometimes hired by the stablecoin issuers themselves, highlighting a conflict of interest. Some audit reports also cover short periods, which is a vulnerability of the auditing governance mechanism, as it becomes easier for the issuer to inflate their performance during this short period. Audit reports also differed in their level of depth, with some reports not fully reviewing Tether’s banking agreements \cite{fernandez2024asset}.}

\journalColorText{Despite these significant limitations, the importance of auditing is empirically shown in \cite{bhambhwani2024auditing}. DeFi protocols with more auditors or higher-quality auditors have higher total value locked (TVL) and higher market capitalization. After the first audit, protocols experience significant increases in value, particularly when the audit is by high-quality auditors. Audited protocols experienced smaller declines in value after the collapse of TerraUSD stablecoin in 2022.}

\insights{\journalColorText{Stablecoin audits show theoretical and empirical value, but also highlight  significant limitations.}}

\journalColorText{\textbf{Methodologies: }Several authors utilize a variety of regression models to analyze the impact of governance-related variables on the performance of the MakerDAO stablecoin platform \cite{zhao2022task, sun2024decentralization, kozhan2022fundamentals}. The explained variables are related to the volatility of the stablecoin and network activity. Explanatory variables include the number of voters, the proportion of large shareholders and Gini coefficient of voting power distribution (measures of voting inequality), the proportion of voters holding a crypt-asset related to the vote, votes controlled by delegates, the type of decision related to the voting task (strategic vs. operational), and information provided to voters measured by the number of webpage links in the voting task's text. The main methodology in \cite{bhambhwani2024auditing} involved using cross-sectional regression analysis to study the impact of audit reports on 316 DeFi platforms, with the explained variable being the platforms' Total Value Locked (TVL) and the explanatory variables related to the number and quality of audits.}

\journalColorText{\citet{moura2022ledger} analyzed how auditors can audit stablecoins by discussing the possible types of transactions. They do not use any empirical data but analyze hypothetical transaction scenarios. For example, if the transaction occurs entirely within a service provider's platform, like an exchange, then it can only be confirmed through the platform's off-chain ledger. For transactions involving transfers to or from a wallet, then on-chain data can be used to verify.}

\journalColorText{\citet{lloyd2023emergent} analyzed voter participation rates within the Curve DeFi platform by examining on-chain data. The study also calculates the cost-per-vote for different methods of acquiring voting weight: direct token locking on Curve, gaining voting weight through Convex (a vote-pooling platform), and bribes through Votium (a platform that uses blockchain to pay users to vote in a certain way).}

\journalColorText{\citet{fernandez2024asset} analyzed audit reports related to Tether in order to assess the quality of the audits. Reports by several firms were analyzed (e.g., Friedman LLP, MOORE, MHA CAYMAN, BDO). The reports were assessed based on several factors: auditor independence, period coverage, and audit scope/depth.}

\textbf{Research Gaps:} Improving stability mechanisms in stablecoins remains a key area of ongoing research. Researchers have proposed interesting ideas to be studied further, such as incorporating algorithmic adjustments to the structure of stablecoins \cite{moin2020sok}. For example, this could include analyzing how stability algorithms perform under different market conditions and developing algorithms that automatically change the mechanisms based on current conditions \cite{moin2020sok}. 

Another critical research area is understanding the effects of different stability mechanisms (e.g., algorithmic vs. custodian-based) and governance structures, such as different voting protocols (e.g., token-weighted or quadratic voting) on the success, stability, efficiency, and equity of stablecoin ecosystems. This research question can be addressed by comparing different ecosystems, as suggested in \cite{zhao2022task}, or the same ecosystems before and after structural changes. It can also be addressed through theoretical modeling and conceptual analysis, potentially drawing analogies from traditional monetary policy.

\citet{charoenwong2023computer} formally demonstrated that purely algorithmic stablecoins cannot achieve provable stability. Building on this, future research could focus on hybrid stablecoins that combine algorithmic mechanisms with collateralization, such as DAI and FRAX, which represent a significant portion of the stablecoin market. Additionally, scenario analysis and stress testing of stability mechanisms are critical for identifying vulnerabilities and ensuring resilience, as emphasized in \cite{kjaer2021empirical}.

\journalColorText{Another important research question to be addressed is how current platforms are governed. This is currently a significant research gap since the majority of stablecoin platforms have not been studied. This relates to how important decisions are made in the ecosystems and what other governance mechanisms are there. Other than examining the platforms' documentation, one innovative method of addressing these questions is through analyzing social media data that can provide some insights about discussion and recommendations relevant to the governance mechanisms. Understanding how to encourage voter participation and prevent malicious activities, such as Sybil attacks \cite{sun2024decentralization}, are also important future research avenues. \citet{zhao2022task} found that votes on operational decisions negatively impact the MakerDAO platform; this opens up a significant research question: how should operational decisions be made, or how can the current decision-making regarding operational decisions be improved?}

\journalColorText{There is a gap in terms of accounting standards surrounding stablecoins, which is a significant topic given that this uncertainty needs to be addressed before stablecoins can be widely adopted by institutions that need to comply with accounting standards. Auditing of stablecoins is also an important topic, given its value in creating confidence. The development of smart, more transparent, and reliable audits is an area of research that is emerging and important \cite{moura2022ledger}. 
}

\subsection{Failure Analysis}
\textit{\textbf{(19) What Are The Causes Of Previous Stablecoin Failures?}}


The TerraLuna collapse was caused by an attack involving a large sell-off, in addition to the excess minting of LUNA. Iron Finance, a stablecoin platform, collapsed due to a situation similar to a bank run. Deleveraging spirals, which occur when automatic liquidations are triggered due to a drop in the stablecoin's collateral, create a downward spiral that amplifies stablecoins' devaluation.

\begin{figure}[htbp]
    \centering
    \includegraphics[width=\columnwidth]{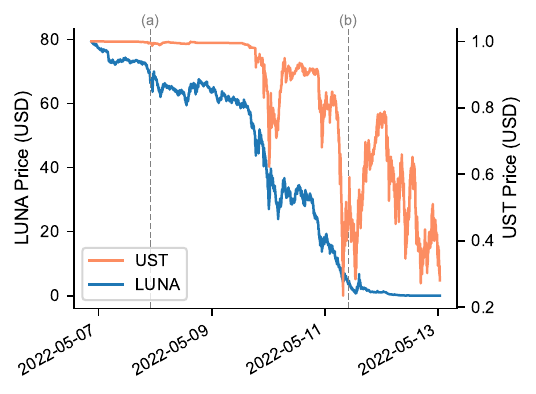}
    \caption{\journalColorText{Minute-by-minute price movements of UST and LUNA during the Terra–Luna collapse in May 2022. Vertical lines indicate two key events: (a) May 7, 22:00: UST first depegs following a liquidity shock on Curve's 3pool; intervention by the Luna Foundation Guard follows. (b) May 11, 10:00: Proposal 1164 is endorsed to increase LUNA minting capacity in an attempt to restore the peg. Data sourced from \href{https://www.binance.com/en/binance-api}{Binance API}}.} 
    \label{fig:luna-ust-minute}
\end{figure}

Four papers discuss the factors that lead to stablecoin failure. 
Two of them analyze the failure of Terra \cite{briola2023anatomy, cho2023token}. At its peak, UST (Terra's algorithmic stablecoin) had a market capitalization of approximately \$18 billion, while LUNA (Terra's native token) had a market capitalization of over \$40 billion. UST maintained its peg to the dollar by offering arbitrage opportunities between LUNA and UST. Terra allowed users to burn \$1 worth of LUNA to mint 1 UST. In this sense, UST was backed by LUNA. Users' trust, speculative investment, high staking rewards, and the large amounts of money backing the Terra platform are all factors that gave LUNA market value. Terra could keep issuing more UST as long as the value of LUNA grew. However, this system was fragile, as a large enough fall in LUNA's market value would mean that there is not enough LUNA to back the \$1 value of all issued UST. It is reported that attackers (with short positions) accumulated a significant amount of UST and suddenly dumped this large amount into a Curve-3pool \cite{briola2023anatomy}. This caused downward pressure on the price, causing a rapid de-pegging of UST. \journalColorText{Fig.~\ref{fig:luna-ust-minute} visualizes this collapse, showing minute-by-minute price movements, of UST and LUNA and marking key events.}

The Terra's protocol began minting more LUNA in an attempt to restore the peg. However, this was a mismanaged policy as it caused hyperinflation of LUNA, leading to further rapid declines. As users had less confidence, selling demand increased. There were not enough assets to back the value of UST. This case highlights the vulnerability of algorithmic stablecoins to extreme market events. 

Drawing parallels with the case of UST, \citet{saengchote2024digital} analyzed the fall of the Iron Finance algorithmic stablecoin. IRON was backed by a combination of USDC and TITAN (the ecosystem's native token). When large TITAN holders began selling their tokens, the price declined. \citet{saengchote2024digital} detected multiple bubble-like patterns in IRON before the collapse. The author argues that algorithmic stablecoins relying on secondary tokens with no collateral are inherently prone to such bubbles, making them vulnerable to collapse when these bubbles subside. \citet{saengchote2024digital} also showed that sophisticated users (large addresses, experienced participants, and those minting rather than buying) exited faster and more profitably.

Further theoretical insights into stablecoin failures are provided by \citet{klages2022while}, who developed a model of deleveraging spirals. Deleveraging spirals happen when stablecoin collateral drops significantly, triggering automatic liquidations of the collateral to prevent under-collateralization. This increases the supply of the collateral in the market, which puts downward pressure on the price of this collateral, causing more liquidations. This concept of deleveraging spirals can also be seen in traditional economics. The 2008 Financial Crisis, where the decline in mortgage-backed securities led to margin calls and massive sales, is a classic example.

\insights{Deleveraging spirals, coordinated selling attacks, and redemption runs (analogous to traditional bank runs) are possible causes of stablecoin failures.}

\textbf{Methodologies:} All of the four papers use a case-study methodology to understand the causes of specific failure cases. \citet{briola2023anatomy} used a network model with cryptocurrencies as nodes and correlations in their price movements as edges. Network metrics such as centrality measures showed that Bitcoin price movements played a role in the collapse of TerraLuna. 
Similarly, \citet{saengchote2024digital} examined blockchain transactions on the Polygon network to study the downfall of Iron Finance. This analysis incorporated data on token minting, redemptions, and swap pool withdrawals. Using the Generalized Supremum Augmented Dickey-Fuller (GSADF) test, the authors identified bubble-like price patterns, while regression models revealed factors driving user behavior during the collapse. In another study, \citet{cho2023token} employed Granger causality tests to see if the value of UST obtainable by swapping it for LUNA influenced the market price of UST, suggesting that under-compensation of UST drove down the market price. Finally, \citet{klages2022while} developed a theoretical stochastic model to demonstrate how leverage and liquidations can trigger stablecoin failures through deleveraging spirals. The model includes collateral value, stablecoin supply, collateralization ratio, and stablecoin price as main parameters. The model shows that negative shocks to collateral value trigger liquidations, pushing the price of the stablecoin to infinity. The model has a key limitation that contributes to this seemingly impractical result. Mainly, the authors ignore the role of arbitrage mechanisms, where arbitrageurs mint new coins (increasing supply) and then sell them in the secondary market for a profit, bringing the price down.

\textit{\textbf{(20) What Are The Impacts Of Previous Stablecoin Failures?}}

Two papers focus their analysis on the impacts of the failure of stablecoins (particularly TerraUSD or UST) \cite{kitzler2023disentangling, lee2023dissecting}. They found that the collapse of TerraUSD (UST) directly caused significant financial losses, harmed several DeFi protocols, spread financial damage across blockchains, and reduced investor confidence in algorithmic stablecoins.

One paper states that the failure of this stablecoin caused the destruction of over \$30 billion in value in just a week \cite{kitzler2023disentangling}. The failure affected multiple other DeFi protocols that were involved with UST. For instance, many liquidity pools on decentralized exchanges like UniSwap and Curve used UST as a primary asset. When UST collapsed, the value of these pools collapsed. Lending protocols like Aave or Compound allowed users to borrow or lend UST \cite{kitzler2023disentangling}. The failure of UST did not only affect the Terra blockchain but also other blockchains that it was bridged to \cite{kitzler2023disentangling}. Moreover, the collapse of UST also caused a crisis of confidence in the entire DeFi ecosystem, as investors lost trust in algorithmic stablecoins. All these impacts demonstrate the systemic risk of DeFi protocols.

\textbf{Methodologies:} \citet{kitzler2023disentangling} used network analysis with nodes representing Ethereum smart contracts or DeFi protocols and edges transactions between these nodes. Using network properties like degree distribution and centrality measures, the authors show the level of interconnectedness of the DeFi ecosystem and how the failure of UST affects the network.

\citet{lee2023dissecting} used the Diebold-Yilmaz (DY) Spillover Framework to study how the crash of Terra-LUNA spilled over to other cryptocurrencies. An extension of the Vector Autoregression (VAR) model, the DY framework has some advantages, including being order-invariant and being able to capture time-varying relationships. The DY framework also more clearly quantifies the overall interdependence in the system. Additionally, the authors used Effective Transfer Entropy (ETE) in a methodology designed to capture information flow between cryptocurrency price time series. Unlike the DY framework, ETE does not assume linearity, is more robust to nonstationarity, and can identify causal relationships. 

\journalColorText{\textit{\textbf{(21) How Can We Better Forecast Stablecoin Failures?}}}

\journalColorText{\citet{ferretti2023using} and \citet{ferretti2024cryptocurrency} investigated the predictive value of Twitter sentiment in forecasting the collapse of two stablecoins, UST and USTC. Their analysis revealed a significant increase in negative tweets preceding the collapse of UST, suggesting that social media can provide potential early warning signals. However, they did not identify a negative correlation between Twitter sentiment and USTC prices, thereby challenging the reliability of social media data in predicting stablecoin crises. Adding to the nuance, their findings also indicated an unexpected surge in positive sentiment during the USTC collapse. This rise in positive tweets may reflect efforts to reassure investors during times of panic or reactions to recovery measures, such as the launch of TerraLuna 2.0, which occurred during the collapse period.}

\journalColorText{\textbf{Methodologies: }\citet{ferretti2023using} used the Twitter API to collect tweets and then filtered them using relevant hashtags and keywords such as \#UST, \#TerraUSD, and \#crypto. This resulted in a dataset of 330,419 tweets from 61,970 unique accounts. A lexicon-based sentiment analysis library, TextBlob, is used to classify tweets based on their polarity (positive, negative, or neutral). The polarity of tweets in the periods surrounding the stablecoin crises is analyzed. In another paper, \citet{ferretti2024cryptocurrency} followed a similar methodology. However, they used a different dataset focusing on USTC instead of UST and consisting of 244,312 tweets from 89,649 unique accounts.}

\textbf{Research Gaps:} Future research ideas proposed by the reviewed papers include analyzing more spillovers related to stablecoin failures \cite{lee2023dissecting, kitzler2023disentangling}, proposing policies or methods to mitigate the risk of failures \cite{cho2023token}, analyzing user behavior during DeFi failures using blockchain transaction data \cite{saengchote2024digital, klages2022while}, and developing better models that can help predict failures \cite{ferretti2024cryptocurrency}.

\citet{klages2022while} have identified deleveraging spirals as a powerful dynamic behind stablecoin collapse. This insight raises an important research question: how can deleveraging spirals in stablecoins be prevented? None of the reviewed papers answer this question.

\journalColorText{The two papers attempting to predict stablecoin failures do so using Twitter data. Using additional variables, such as stablecoin network activity or news articles, can offer additional insights. Utilizing other forms of social media data can also show value. Future research can also look at whether a subset of social media accounts has higher predictive value; for instance, it is possible that accounts with more influence are able to predict crises to a greater extent than other accounts. The two papers also use a lexicon-based sentiment analysis technique (TextBlob). This technique might not be the most suitable way to classify the sentiment of finance-related tweets. Lexicon-based methods can struggle to detect more complex patterns of language compared to more sophisticated machine learning models. Previous research has demonstrated the value of using context-specific machine learning models for financial tweets sentiment analysis \cite{mahrous2023role}. Models taking into account the context of financial tweets, such as emojis used for financial discussions, can be developed and utilized in future research (e.g., \cite{FinTwitBERT}.}

Moreover, the majority of papers studying stablecoin failures focused on the TerraLuna case, leaving a need to examine other cases of stablecoin failures (e.g., Basis Cash, NuBits, Iron Finance, BeanStalk, Deus Finance, flexUSD, Acala USD).

\subsection{Security And Privacy}

\textit{\textbf{(22) What Are The Privacy Concerns Related To CBDCs And How Are They Being Addressed?}}

Only one paper focuses on the privacy of stablecoins. \citet{tronnier2021privacy} analyzed how 78 different central banks address privacy in their CBDC-related publications and pilot projects, uncovering several valuable findings.

The analysis revealed that most central banks either do not address privacy or discuss it only superficially. However, some central banks prioritize privacy more extensively, with the most notable being the Bank of Canada, the Bank of Italy, the European Central Bank (ECB), and the Monetary Authority of Singapore. Other central banks, such as those of France, Thailand, and Denmark, also engage in more detailed discussions on privacy.

A common challenge highlighted is the trade-off between central bank oversight and user privacy. For example, Know-Your-Customer (KYC) procedures, required by central banks for regulatory compliance, inherently reduce privacy.

To address such concerns, central banks proposed various privacy-enhancing mechanisms. Specific to Europe, the General Data Protection Regulation (GDPR) defines transaction data as personal data and restricts its processing. Central banks, including the Norges Bank (Norway), argue that GDPR compliance must be integrated into CBDC design, significantly enhancing transaction privacy. Other proposed measures include ensuring that user identity data is not published on the blockchain and offering users cryptographically signed IDs.


Several central banks propose mechanisms for anonymous payments. For example, the European Central Bank (ECB) suggests using anonymous vouchers, which can be purchased from third-party intermediaries to enable small anonymous CBDC transactions. Privacy-enhancing technologies such as zero-knowledge proofs and differential privacy, highlighted by the Bank of Canada, offer additional tools to safeguard user privacy. Other techniques, such as zero-knowledge proofs and differential privacy, both highlighted by the Bank of Canada, offer additional tools to protect user privacy. To prevent long-term tracking of users' transactions, rotating public keys is mentioned as a technique to increase anonymity.

The Dutch Central Bank proposes separating transaction data from user balance data, limiting access to intermediaries. The ECB pilot project discusses chain snipping as a method to reset transaction history, preventing the creation of long-term knowledge graphs of users' payment activity. Additional privacy-enhancing methods, including multi-party computation, enclave computing, and permissioned distributed ledger technology, are also mentioned by several central banks.

\textbf{Methodologies:} The methodology of 
\citet{tronnier2021privacy} conducted a systematic literature review of CBDC-related publications by 78 central banks. The study systematically searched each central bank's website for relevant documents and supplemented this with Google searches using keywords such as "CBDC," "Central Bank Digital Currency," and the name or country of the central bank. Only complete, English-language publications that directly addressed CBDCs and privacy considerations were included in the analysis. The authors categorized central banks based on how extensively they covered privacy, focusing exclusively on retail CBDCs while excluding wholesale CBDCs intended for interbank settlements.

\textit{\textbf{(23) What Are The Security Threats Related To CBDCs And Stablecoins?}}

Only two papers focus on security threats related to CBDCs or stablecoins. 

\citet{hans2023blockchain} identified 39 CBDC security threats using the STRIDE modeling framework. Among these, more threats were related to Spoofing (7), Tampering (10), and Information Disclosure (12), while fewer were associated with Denial of Service (4), Repudiation (2), and Elevation of Privilege (4). 

Many threats identified were general cybersecurity threats, not specific to CBDCs or distributed ledger technology (DLT). These include attackers impersonating legitimate users in customer support interactions, introducing backdoors by tampering with hardware during chip manufacturing and deploying Trojan malware or SQL injections. Other general examples are supply chain attacks, SMS interception, network eavesdropping, server flooding to disable access, manipulating access policies for privilege escalation, and exploiting vulnerabilities in user interfaces to gain admin access.

However, several threats were specific to CBDCs and DLT. These include attackers masquerading as legitimate nodes using compromised certificates, exploiting cryptographic vulnerabilities (e.g., collision attacks), falsifying transaction records within the ledger, and targeting intermediaries like miners, financial institutions, and oracles. Additionally, sending false transactions to overwhelm the consensus mechanism poses a unique threat to DLT-based systems.

\citet{mell2023understanding} highlighted additional security threats associated with Central Bank Digital Currencies (CBDCs). These include the unauthorized creation of stablecoins caused by software defects and the theft of on-chain collateral due to vulnerabilities in smart contract code. Other risks involve malicious updates to smart contracts. Furthermore, attackers gaining control of a majority of mining hardware or staked funds can exploit the system.

\textbf{Methodologies:} \citet{hans2023blockchain} employed threat modeling using the STRIDE methodology after creating an abstract high-level architecture of a CBDC. STRIDE classifies threats into six categories: Spoofing, Tampering, Repudiation, Information Disclosure, Denial of Service, and Elevation of Privilege \cite{khan2017stride}. The abstracted system includes key entities (e.g., central banks, financial institutions, and end-users) and their interactions.  Specific use cases for this abstract CBDC were developed, illustrating workflows for transactions and access controls. Sequence diagrams illustrate the interactions among entities in both permissioned and permissionless blockchain environments, as well as in account-based and token-based access scenarios. Threats were identified by analyzing the abstracted system architecture and the modeled use cases.

In contrast, \citet{mell2023understanding} consulted a range of research papers and reports from reputable sources, such as the Federal Reserve System and the U.S. Treasury, to identify security threats. However, the specific methodology they used for this process is not explained.

\textit{\textbf{(24) How Can Market Manipulation Related To Stablecoins Be Detected?}}

One paper focuses on detecting market manipulation related to stablecoins. \citet{zhu2024data} introduced an efficient method for detecting market manipulators in blockchain-based networks, including stablecoins. The method detects traders who played roles in destabilizing the Terra stablecoin. 

\textbf{Methodologies:} 
The market manipulation detection method presented in \cite{zhu2024data} centers on identifying nodes in the network that have a significant but short-lived influence, with the argument being that such nodes are more likely to be manipulators. The authors rank nodes using a scoring method called Node Frequency-Inverse Appearance Frequency (NF-IAF). This method is inspired by Term Frequency-Inverse Document Frequency (TF-IDF), a widely used metric in information retrieval. NF measures how frequently a node participates in a specific motif relative to all nodes participating in that motif. IAF quantifies how rare a node's participation in a motif is across all days; rare appearances are more significant. Motifs, in this context, are recurring patterns of interaction between three nodes in the network. Only 5 of the 16 possible three-node motif patterns are included in the analysis. These include motifs where a central node receives/sends transactions from/to two other nodes and the wash-trading or money-laundering motif with a circular trading pattern (A to B to C to A).

\textbf{Research Gaps:} Research on the security and privacy of stablecoins and CBDCs is surprisingly limited, with only four relevant papers identified in our literature search. This is despite the security and privacy of stablecoins and CBDCs being a significant issue for many reasons, including their impacts on user trust, protection against hacks and manipulation, ethical considerations, and regulatory compliance. This gap presents significant opportunities for future research.

Potential areas of investigation include analyzing methods to enhance privacy in stablecoins and CBDCs, measuring user demand for privacy, assessing the trade-offs between privacy and factors such as efficiency, transparency, scalability, and regulatory control, and designing and implementing privacy-focused stablecoins. In this regard, the only contribution is \cite{tronnier2021privacy}. This study analyzes publications by central banks to examine their discussions about privacy. Although interesting, this work excluded any non-English publication, which should be considered in future research, neglecting the position of non-English-speaking countries regarding end-user privacy in CBDCs.

Security breaches can result not only in the possible loss of money but also in the destabilization of entire economies, particularly in the case of CBDCs. Stablecoins are interconnected with the entire DeFi ecosystem and potentially with traditional finance, increasing the systemic risk of security breaches. \citet{hans2023blockchain} identified many security threats using the STRIDE methodology. Other researchers can use different methodologies, such as attack tree analysis or analysis of stablecoin smart contract code. 
Moreover, \citet{hans2023blockchain} performed the analysis on a general abstraction of stablecoins. Future research can analyze actual projects and provide a comparative analysis of their security. This can help investors be aware of which stablecoins are more or less secure. The usage of AI for both attacking and defending stablecoin security is also a promising area of research.

Manipulation specific to stablecoins is also a scarcely researched topic that warrants more focus by researchers. \citet{zhu2024data} detected different patterns of manipulation resembling coordinated buying and selling during the TerraLuna collapse---a phenomenon well documented in broader cryptocurrency markets \cite{charfeddine2024drives}. \citet{routledge2022currency} developed a theoretical model of speculative attack manipulation in stablecoins and proposes a solution to prevent it. Some papers also study the manipulation of cryptocurrencies through minting stablecoins \cite{griffin2020bitcoin, kristoufek2021tethered}. This leaves much room for future research to develop more advanced manipulation detection techniques, theoretically model different types of manipulation, and propose methods to combat manipulation. Types of manipulation that may exist in the stablecoin ecosystem but remain unexplored include oracle manipulation attacks, voting and governance manipulation (e.g., the BeanStalk stablecoin case), and the exploitation of vulnerabilities in smart contract mechanisms (e.g., the Acala Dolar stablecoin case). Moreover, none of the reviewed papers shed light on the measures existing stablecoins take to combat manipulation. This can be done in future research by analyzing technical documentation related to the stablecoins.

\journalColorText{\section{Position and Policy Papers}
\label{sec:policy_analysis}}

\journalColorText{\subsection{Position Papers Providing an Overview of Stablecoins and their Implications}}

\journalColorText{Beyond their economic and technological functions, stablecoins and CBDCs intersect with core questions of geopolitics, social impact, public policy, and regulatory governance. This section analyzes academic position and policy papers that explore these dimensions. The first sub-category examines the general characteristics and implications of stablecoins and CBDCs, focusing on their potential to impact financial systems, traditional monetary policy, and geopolitics. The second sub-category addresses legal and regulatory issues, including cross-border laws, consumer protection, reserve transparency enforcement, and the risks stablecoins pose to national monetary sovereignty. Collectively, these papers highlight ideological, institutional, and regulatory concepts that shape the global discourse on stablecoins and CBDCs.}

\journalColorText{\textit{\textbf{(25) What Are The General Characteristics And Implications Of Stablecoins And CBDCs?}}}

\journalColorText{25 position papers discuss stablecoins and CBDCs in general, highlighting their general characteristics, potential, and broader implications. The main results of this section are the following: stablecoins are growing because they combine fiat-currency stability with blockchain advantages like decentralization, smart contract programmability, and DeFi platforms interoperability; in order to counterbalance private stablecoins, many central banks aim to retain control by issuing CBDCs; private stablecoins and CBDCs compete against each other, as well as with traditional methods like bank deposits, but can coexist; CBDCs issued by different countries compete against each other and can be used as geopolitical tools; stablecoins offer promise to bank the unbanked; cooperation and interoperability between different CBDCs are highly useful; CBDCs threaten old monetary policy tools but offer potentially novel tools for central banks, like specifying what certain money can be spent on.}

\journalColorText{Stablecoins can be broadly categorized into four types: fiat-collateralized, crypto-collateralized, commodity-collateralized, and algorithmic \cite{jayapal2023insight}. The different methods of collateralization and algorithmic stabilization mechanisms are discussed in more detail in research question 17. 
Stablecoins backed by off-chain assets, such as fiat or commodities, typically require independent auditing to increase trust.}

\begin{figure}[htbp]
     \centering
     \includegraphics[width=0.48\textwidth]{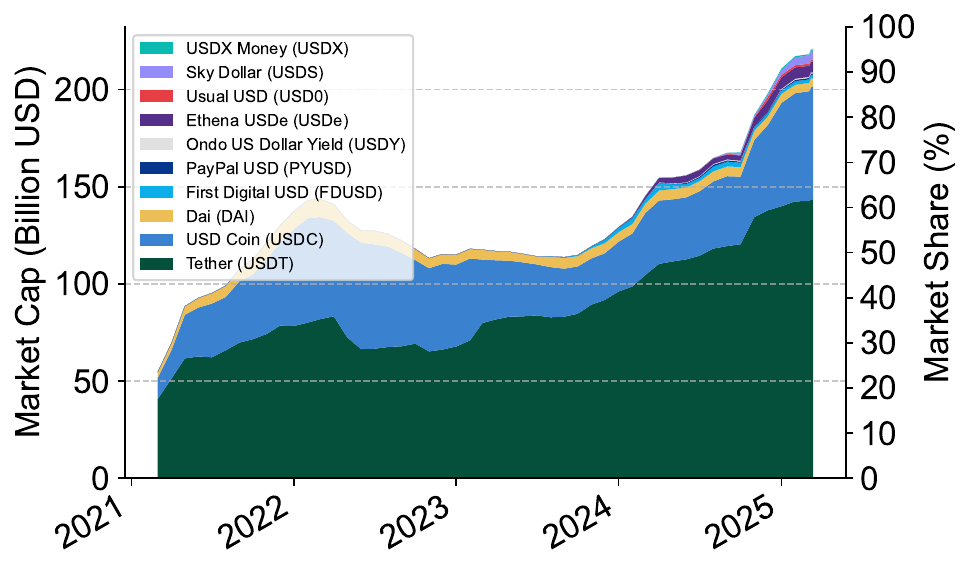} 
     \caption{\journalColorText{Market capitalization and market share of major stablecoins from 2021 to 2025. Data was sourced from \href{https://www.coingecko.com/}{CoinGecko API}.}}
     \label{fig:marketcap}
 \end{figure}

\journalColorText{The stablecoin market is characterized by high concentration and USD dominance. Fig.~\ref{fig:marketcap} highlights the dominance of Tether (USDT) and USD Coin (USDC), while also showing the gradual emergence of new entrants such as PayPal USD (PYUSD), Ethena USDe, and First Digital USD (FDUSD). Notably, the stablecoins in Fig. \ref{fig:marketcap}, which together represent more than 95\% of market cap, are all pegged to the US dollar and not any other national currency.}

\begin{figure}[htbp]
    \centering
    \includegraphics[width=0.95\columnwidth]{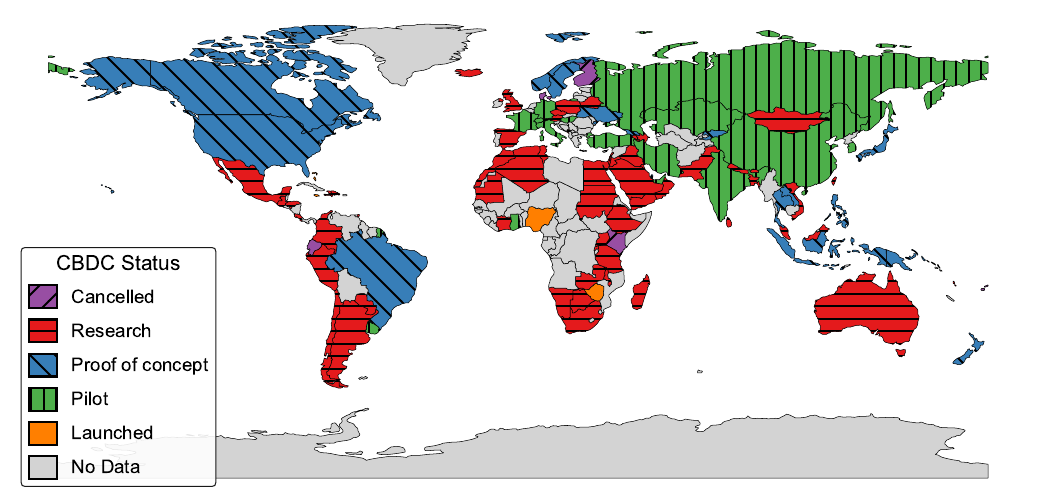}
    \caption{\journalColorText{Global status of Central Bank Digital Currency (CBDC) projects. Countries are color-coded by project stage. Data was sourced from \href{https://cbdctracker.org/.}{CBDC Tracker}}.}
    \label{fig:cbdc_world_map}
\end{figure}

\journalColorText{The rise of stablecoins and the growing interest in CBDCs are noticeable trends. The demand for stablecoins is driven by their ability to combine the stability of fiat currencies with the advantages of blockchain-based currencies, such as faster, more secure transactions and smart-contract programmability  \cite{hull2024properties}. In response to the growth of private stablecoins such as Tether, central banks are interested in developing CBDCs to maintain control over national financial systems \cite{copeland2020global}. Many countries have already shown significant interest in CBDC technologies, with 3 launching a CBDC (Nigeria, the Bahamas, Jamaica), 44 launching pilot projects, and many more investing in related research, as shown in Fig.~\ref{fig:cbdc_world_map} \cite{atlanticcouncil_cbdc_tracker}. Stablecoins and CBDCs compete against each other \cite{adrian2021rise}; however, some researchers argue that it is best that they coexist \cite{bolt2022getting, zatti2022economic, cesaratto2023central}. Digital currencies also threaten to shift money away from traditional bank deposits \cite{bindseil2019central}; \citet{castren2022digital} argued that this issue is difficult to solve. In addition, CBDCs involve a dynamic of competition between countries. China’s digital yuan and other CBDC projects are positioned as strategic tools to reduce dependency on the U.S. dollar and enhance geopolitical standing \cite{allen2022fintech, wang2023coming}. USD-backed stablecoins, on the other hand, could potentially lead to de facto dollarization in economically smaller nations \cite{li2021potential}.}

\insights{\journalColorText{CBDCs compete on a geopolitical level, with some nations seeing them as a way to reduce reliance on the U.S. dollar.}}

\insights{\journalColorText{USD-backed stablecoins, representing more than 99\% of stablecoins in circulation, could lead to de facto dollarization.}}


\journalColorText{Stablecoins promise several advantages over traditional fiat currencies, including more efficient cross-border transfers \cite{prasad2023will}, programmability, and enhanced traceability. Stablecoins could provide a cheaper and more accessible alternative for the large unbanked population. According to the World Bank, around 1.4 billion adults globally remain unbanked as of 2021 \cite{worldbank2021unbanked}. These unbanked individuals could potentially access decentralized finance (DeFi) platforms if they simply have a mobile phone and the required knowledge \cite{bagis2022digital}. Stablecoins also allow for interaction with DeFi platforms and smart contracts, which creates significant opportunities for innovation. Examples include automated payments, usage with IoT devices, automated trading, decentralized lending and borrowing, investment in tokens, etc. Moreover, CBDCs are seen as tools for central banks to mitigate issues of moral hazard by limiting their ability to excessively print money, fostering more confidence in central banks. Such conditions can be programmed in the CBDCs. Additionally, CBDCs could equip central banks with new tools for monetary policy \cite{bagis2022digital, bindseil2019central}. For example, central banks could program money to automatically and transparently issue stimulus funds. Welfare payments could be restricted to specific categories of spending. Money could expire if not used. CBDCs could also be used to apply different interest rates to different segments of society, for instance. Stimulus payments could be made directly to consumers instead of going through banks.}

\insights{\journalColorText{Stablecoins are gaining traction due to their ability to combine fiat currency stability with blockchain advantages such as decentralization, programmability, and interoperability with DeFi platforms.}}

\journalColorText{\textbf{Methodologies: }The methodologies of this section are primarily based on literature reviews and qualitative analyses. Some studies use qualitative assessments of historical monetary shifts \cite{bindseil2019central} and the economic roles of traditional versus digital currencies \cite{bordo2021central}. Many others analyze policy and legal frameworks, primarily to explore the position of central banks in the current ecosystem \cite{wang2023coming}. Others qualitatively discuss technical concepts related to stablecoins and CBDCs, such as latency and scalability \cite{hull2024properties}. These discussions often do not follow a systematic methodology or rigorous theoretical or empirical analysis but aim to introduce the concepts to a non-technical audience.}

\journalColorText{Some theoretical concepts (e.g., Fisher equation, the quantity theory of money, endogenous money supply, social positioning theory, and modern monetary theory) are used to simulate specific impacts of stablecoins and CBDCs on monetary policy and financial stability \cite{li2021potential, bagis2022digital, morgan2022systemic}}. 

\journalColorText{Papers in this section also often employ case studies to explore previously implemented CBDC and stablecoin projects, assessing their properties and challenges.}

\journalColorText{\textbf{Research Gaps: }
Future research topics regarding the characteristics and implications of stablecoins and CBDCs span a broad range of regulatory, social, economic, technological, and geopolitical areas. A recurring theme is the need for global regulatory harmonization; future research can explore frameworks to manage international stablecoin regulation aiming to prevent regulatory arbitrage and address jurisdictional discrepancies \cite{adrian2021rise, copeland2020global, hofmann2020changing}. This is also discussed in research question 26.}

\journalColorText{Another important area for future research relates to financial inclusion, helping digital money reach underserved populations \cite{allen2022fintech, temperini2023democratizing}. Future research can investigate the barriers to the adoption of digital currencies in regions with limited financial inclusion and propose innovative solutions, such as offline-compatible CBDCs or evidence-based user interface designs.} 

\journalColorText{On the geopolitical front, the implications for currency sovereignty and dollar hegemony in light of CBDCs are mentioned as future research avenues \cite{cesaratto2023central, fantacci2024stablecoins}. Future work can analyze the impacts of adopting USD-backed stablecoins on smaller economies.}

\journalColorText{\citet{dowd2024so} highlighted that previously implemented CBDC projects have largely failed. This includes CBDCs implemented by Finland, Ecuador, The Bahamas, the East Caribbean, Jamaica, China, and Nigeria. While some possible reasons for the failures have been discussed in \cite{dowd2024so}, no rigorous empirical analysis has been done in this regard. Future work can give more attention to this critical topic.}

\journalColorText{Moreover, more studies are needed on monetary policy adaptations to CBDCs, addressing the impacts of CBDCs on traditional monetary policy tools and the opportunities for novel monetary policy tools using CBDCs \cite{bagis2022digital}.}

\journalColorText{\subsection{Legal and Regulatory}
\textit{\textbf{(26) What Are The Regulatory Challenges Related To Stablecoins And CBDCs, How Have Regulators Responded, And How Should These Challenges Be Dealt With?}}}

\journalColorText{The main results of this section are the following: there is a need for cross-border regulatory cooperation to combat regulatory arbitrage and international crimes related to stablecoins; to protect consumers, regulators need to ensure quality in stablecoin reserves and safeguard consumer data; the legal treatment within existing financial reporting standards remains ambiguous; regulators, particularly those overseeing smaller economies, face a challenge of maintaining monetary sovereignty against stablecoins; Facebook's Libra stablecoin project faced increased scrutiny from regulators and ultimately ended up winding down.}


\journalColorText{Stablecoins can be issued by an entity based in one country but are easily accessible worldwide. With transactions occurring globally, disputes involving stablecoins often require cooperation between regulators \cite{schwarcz2022regulating}. Cooperation between regulators is also necessary to mitigate regulatory arbitrage, where firms choose to register in countries offering more lenient regulations. International crimes such as sanctions evasion \cite{travkina2022stablecoin}, money-laundering, and terrorism financing \cite{ferreira2021curious} can be facilitated using stablecoins; this presents a challenge to regulators as they decide what regulations can be implemented to address such crimes. \citet{schwarcz2022regulating} proposed that researchers develop a model-law strategy: a set of principles or rules that guide regulators and possibly lead to more regulatory harmonization.}

\insights{\journalColorText{Stablecoins, being global assets, require cross-border regulatory cooperation to reduce regulatory arbitrage and combat international crime.}}

\journalColorText{Consumer protection laws are necessary in order to prevent malicious behavior by stablecoin issuers, including mismanagement of reserves or consumer data \cite{shi2021legislation}. This includes enforcing reserve quality and transparency measures by regulators \cite{shi2021legislation}. The categorization of stablecoins within existing financial reporting standards also remains unclear \cite{fernandez2020stable, hampl2021can}. \citet{hampl2021can} argued that the low volatility of some stablecoins allows them to be qualified as cash-equivalent under International Financial Reporting Standards (IFRS). However, their low volatility does not necessarily mean stablecoins meet the criteria for formal classification as cash-equivalent. \citet{fernandez2020stable} showed that current Ethereum token standards, including ERC20, ERC721, and ERC1155, do not meet regulatory requirements or standards like ISO 20022, which is essential for adoption in institutional payments. Newer standards like ERC1450 and ERC2020 are more compliant but still in development.}

\insights{\journalColorText{The classification of stablecoins under legal and accounting standards remains ambiguous, hindering adoption.}}

\journalColorText{Another main challenge facing regulators is how to set up regulations to protect monetary sovereignty since stablecoins have the potential of replacing national currencies, particularly for smaller countries or those with weak local currencies \cite{arauz2021international}.}

\journalColorText{\citet{zetzsche2021regulating, read2020libra} reported the heightened scrutiny regulators showed towards the Libra stablecoin project. 
Central banks, financial regulators, and international organizations like the G7, Financial Stability Board (FSB), and the Financial Action Task Force (FATF) emphasized the risks posed by Libra to the current financial system. In the United States, both the Federal Reserve and Congress expressed serious concerns about Libra’s impact on the dollar’s role as a global reserve currency and its potential to operate outside traditional regulatory frameworks. Similarly, European regulators, including the European Central Bank and the Bank of England, stressed that Libra would need to meet the highest regulatory standards. France and Germany went so far as to block the project entirely. Facebook's Libra project was announced in 2019, but to this day has not been delivered, serving as a case for the regulatory complexities related to stablecoin projects.}

\insights{\journalColorText{The Libra case highlights regulatory pushback against private stablecoin projects.}}

\journalColorText{\textbf{Methodologies: }The majority of papers performed a qualitative or comparative regulatory analysis by reviewing regulatory documents \cite{zetzsche2021regulating, ferreira2021curious, read2020libra, travkina2022stablecoin, shi2021legislation, ikeno2022soundness, schwarcz2022regulating}. This methodology aims to evaluate existing regulatory frameworks across jurisdictions, which gives an idea of the challenges related to stablecoins and CBDCs regulators are trying to address and their responses to these challenges. It also gives researchers an idea of the different regulatory approaches used by the different jurisdictions. Analyzing current regulatory frameworks also serves researchers who want to spot gaps in the current frameworks and propose ways to address these gaps.}

\journalColorText{Other than regulatory documents, the following documents are used to gain an understanding of the regulatory concerns related to stablecoins or CBDCs: projects' white-papers \cite{zetzsche2021regulating}, statements by international organizations such as the G7 \cite{read2020libra}, news articles \cite{ferreira2021curious}, and academic literature \cite{travkina2022stablecoin}.}

\journalColorText{Some papers draw parallels between the stablecoins/CBDCs and traditional financial assets in an attempt to argue for or against similar regulatory treatments. \citet{hampl2021can} analyzed the volatility of stablecoins and argued that the majority of stablecoins are stable enough to be considered cash or cash equivalents under IFRS standards. Several papers also compare stablecoins to money market funds \cite{zetzsche2021regulating, eichengreen2022stablecoins, read2020libra}.}



\journalColorText{\textbf{Research Gaps: }}\journalColorText{Important future research topics mentioned include global regulatory harmonization and cooperation, consumer protection concerns, monetary policy sovereignty, sanction evasion using stablecoins, and geopolitical implications.}

\journalColorText{Stablecoins operate across borders, challenging current national-focused financial regulations. Harmonized regulatory frameworks are essential in reducing regulatory arbitrage, where stablecoin entities choose the most lenient jurisdictions \cite{zetzsche2021regulating}. This pushes countries on a slope where they are all pressured to offer more lenient regulations to attract money related to stablecoin entities, even if that means harm to the markets. Researchers can help by providing a comparative analysis of the current regulatory landscape as a guide to regulators. They can also propose global coordination models \cite{schwarcz2022regulating}. They can also raise awareness of regulatory arbitrage risks through economic models or empirical analysis.}

\journalColorText{Consumer protection in the stablecoin ecosystem has become a significant concern for regulators. Stablecoin entities are trusted by consumers for maintaining stability or being a custodian of reserves \cite{shi2021legislation, ferreira2021curious}. However, many potential users of stablecoins belong to financially marginalized populations, making them particularly vulnerable. Compounding this issue, stablecoins often operate outside the control of regulatory frameworks, especially given the ability to practice regulatory arbitrage. Moreover, stablecoin entities control vast sums of money, including investments in cryptocurrencies, that may be inaccessible to regulators.}

\journalColorText{These factors present unique challenges to consumer protection in the stablecoin space. Researchers can propose enforceable regulatory frameworks and mechanisms for consumer recourse when disputes or failures occur. Researchers can also shed light on cases where consumers were treated fraudulently to highlight the importance of consumer protection. Designing evidence-based awareness programs can also help consumers protect themselves.}


\journalColorText{Another area of future research involves how stablecoins could be used to circumvent economic sanctions \cite{travkina2022stablecoin}. Future research can measure the current sanctions-evading behavior by analyzing stablecoin transactions. Researchers can also propose countermeasures regulators can take to mitigate such behavior.}

\journalColorText{According to \citet{hampl2021can}, the low volatility of certain stablecoins suggests their potential classification as cash equivalents under International Financial Reporting Standards (IFRS). However, low volatility alone is insufficient to fully satisfy the requirements for official cash-equivalent status. Further research is needed to identify the specific criteria stablecoins must meet to achieve this classification and to explore whether alternative classifications might be more appropriate.}

\section{Discussion}
\label{sec:discussion}

In this section, \journalColorText{we first discuss the research trends within the three macro-categories considered in this study, highlighting the areas that have been extensively investigated and identifying research questions that remain under-explored. Next, we address additional research gaps observed during our analysis, including research questions with conflicting results from different contributions. Finally, we discuss the various data types and sources commonly used in this research domain.}
\journalColorText{\subsection{Research Trends}
Some research questions have received significantly more attention than others, as shown by Table~\ref{tab:categories}. Fundamental topics, such as the extent of stablecoins' stability and the mechanisms that make them stable, naturally attract greater focus. Additionally, certain research questions are easier to address, as they require less specialized knowledge, rely on more accessible data, or involve less time-consuming methodologies. For example, the research question related to the correlation between stablecoin prices and the prices of other assets is the focus of more than 30\% of the papers analyzed in this study. This question can be addressed using price data, which is easily accessible through data aggregators or exchanges. While daily price data from sources like CoinMarketCap is widely used due to its accessibility, it is often less valuable than higher-granularity price data (e.g., minute-by-minute) that is usually not available on public datasets and must be self-collected by researchers. This issue is discussed in more detail in Subsection~\ref{subsec:datatypes}.}

\journalColorText{In contrast, other research questions have received attention from only a few papers or, in some cases, just one, highlighting the potential existence of research gaps in these areas. For example, only two papers analyze security concerns related to stablecoins, and they do so in a broad and conceptual manner without analyzing specific stablecoins. Only one paper focuses on privacy issues, specifically discussing how central banks address privacy in their publications \cite{tronnier2021privacy}. This paper adequately addresses the research question regarding the importance central banks assign to privacy in CBDCs. It does so by systematically analyzing documents from 78 central banks and categorizing them based on the emphasis they place on privacy in CBDCs. However, several other research questions related to privacy in stablecoins remain unexplored. These include the implementation of privacy-preserving mechanisms in stablecoins and the current levels of privacy in the existing stablecoins.}

\journalColorText{Similarly, the governance mechanisms of stablecoins---unlike their stability mechanisms---have received relatively limited attention. Only two governance mechanisms are studied: voting and auditing. Moreover, research on governance mechanisms focuses mainly on one of two stablecoin platforms: MakerDAO and Curve Finance. Studies on MakerDAO have revealed that voting power is highly concentrated among a few large token holders \cite{zhao2022task, sun2024decentralization, kozhan2022fundamentals}. This concentration poses risks to the platform's decision-making, and future research should explore mechanisms to mitigate these risks, such as alternative voting structures. Additionally, governance through voting is susceptible to manipulation via bribery marketplaces like Votium. \citet{lloyd2023emergent} demonstrates that actors aiming to influence decisions on Curve Finance can do so more efficiently by purchasing votes through Votium than by directly acquiring governance tokens from the platform itself. As for auditing, \citet{bhambhwani2024auditing} showed the positive impact of audits on DeFi platform value, while \citet{fernandez2024asset} and \citet{moura2022ledger} highlighted the limitations of current stablecoin auditing practices, such as inconsistencies, conflicts of interest, and limited scope, emphasizing the need to develop higher quality stablecoin audit standards.}

\journalColorText{Furthermore, only four papers investigate the causes and impacts of stablecoin failures, and they focus exclusively on two cases: TerraLuna and IRON \cite{briola2023anatomy, klages2022while, saengchote2024digital, cho2023token}. By analyzing blockchain transaction data, they identify underlying causes, including liquidity pool attacks, misaligned incentives (a flawed stablecoin redemption mechanism), and deleveraging spirals. They also quantify the impacts of the failures to the magnitude of tens of billions of USD. Numerous other stablecoin failures have been documented in the grey literature but remain understudied, including Basis Cash, NuBits, Iron Finance, BeanStalk, Deus Finance, flexUSD, and Acala USD. Different projects use distinct designs; their analysis can reveal novel common failure risk factors for stablecoins. For example, many sources in the grey literature reported that several stablecoins have failed due to smart-contract exploits \cite{ligon2022acala}, governance manipulation using flash loans \cite{immunefi2023beanstalk}, and regulatory constraints \cite{dale2018basis}. Studying more cases also strengthens the generalizability of stablecoin failure reasons. It also sheds light on the mechanisms that failed to prevent failure.}

\journalColorText{A few papers discuss market manipulation related to stablecoins. The techniques and attacks studied focus on one of two mechanisms. The first is minting unbacked stablecoins to inflate the prices of other cryptocurrencies \cite{griffin2020bitcoin}. The second is related to speculative attacks where attackers accumulate large amounts of stablecoin and suddenly sell them in an attempt to de-peg the stablecoin and benefit from short positions \cite{routledge2022currency, briola2023anatomy, saengchote2024digital}. Building a taxonomy of manipulation techniques relevant to stablecoins and records of previous manipulation cases, along with their impacts, can be a promising future research direction. Investigating how to mitigate the risk of manipulation is also an active area of research.}

\subsection{Other Research Gaps and Conflicting Results}
Several research gaps persist also within the research questions that received greater attention. For example, out of the 37 papers studying the relationship of markets of different assets, only one paper analyzes the relationship of stablecoins with respect to each other. Moreover, the stability of some important stablecoins has not been examined, with the majority of papers focusing on USDT, USDC, and DAI only. The second research question includes papers studying the impact of different environmental factors on stablecoin markets. However, some factors are understudied (e.g., interest rates, news announcements, and regulatory factors), and some potentially impactful aspects are not studied at all (e.g., activity on different DeFi platforms, emerging market activity and announcements of stablecoin issuers).

There are also research questions with conflicting results, possibly requiring more robust research to settle the debates. 
One example is related to the relationship between the prices of stablecoins and the value of Bitcoin. For instance, \citet{hatem2022roles} and \citet{chen2022volatility} reached opposing results regarding the impact of Bitcoin price movements on stablecoins. On the one hand, \citet{hatem2022roles} applied a bivariate VARMA-BEKK-GARCH model on BTC/stablecoin pairs from October 8, 2018, to August 17, 2020, and found that BTC price movements impact stablecoins' prices and trading volumes. \journalColorText{On the other hand, \citet{chen2022volatility} applied multivariate GARCH and VAR models on the prices of Bitcoin and five stablecoins considering a sensibly longer time window, that is, from July 1, 2014, to February 15, 2022, finding that BTC price movements do not impact the stablecoins' prices.} The likely explanation lies in the sensitivity of the econometric methods used in these studies to factors such as model specifications, data granularity, and sample periods, all aspects requiring further investigation.

There is also a conflict regarding whether stablecoin issuance is used to manipulate other cryptocurrencies.
\citet{griffin2020bitcoin}, for example, argued that Tether issuances were used to inflate Bitcoin prices, while \citet{wei2018impact} found no evidence of such manipulation. \citet{griffin2020bitcoin} used clustering and regression models on highly granular intraday transaction-level blockchain data, while \citet{wei2018impact} used Granger Causality and VAR models on daily price data. When evaluating both papers, \cite{griffin2020bitcoin} appeared to have a stronger methodology for addressing the topic of Tether manipulation of Bitcoin prices. This is because \citet{griffin2020bitcoin} used more granular data, allowing the precise timing of the impacts of Tether minting transactions on Bitcoin prices and trading volume. In general, markets react very quickly to new substantial information, and the daily granularity used in \cite{wei2018impact} can mask these quick reactions. \citet{griffin2020bitcoin} did not just examine the correlation of prices but used on-chain data to trace Tether flows through particular exchanges and identified the specific accounts using the minted Tether to prop up Bitcoin prices and trading volumes. This allowed them to offer a multifaceted analysis that approaches the problem from different angles and supports conclusions with a more robust methodological framework. 

Another area of conflict relates to the stability of algorithmic stablecoins. While several papers propose various algorithmic stablecoin designs (e.g., \cite{dong2019elasticoin, heinonen2021creation}), their theoretical instability \cite{charoenwong2023computer}, and empirical volatility have been highlighted \cite{jarno2021does}. This conflict arises, in part, from the theoretical impossibility of guaranteeing absolute stability \cite{charoenwong2023computer}, contrasted with the practical efforts to achieve relative stability through algorithmic mechanisms. However, such theoretical limitations do not entirely rule out the possibility of achieving practical, albeit imperfect, stabilization. Furthermore, the stability of algorithmic stablecoins can vary across different designs. The theoretical instability of stablecoins has been demonstrated in a single paper \cite{charoenwong2023computer}, focusing solely on a purely algorithmic design.

There is also mixed evidence on the stability of stablecoins during crises. Different authors achieved different results depending on the specific stablecoins and the particular crises studied. In some crises, some investors may prefer traditional (supposedly) safer assets like government-backed securities over stablecoins, leading to outflows from the stablecoin market \cite{oefele2024flight}. This was evident during the SVB banking crisis, as stablecoins held a significant amount of reserves with failed banks. The FTX exchange crisis, on the other hand, prompted investors to seek refuge in stablecoins \cite{galati2024market}. Different stablecoins are also impacted differently during the same crisis. For instance, during the Silicon Valley Bank crisis,  
the stablecoins USDC, DAI, and TrueUSD dropped significantly below the peg, while USDT and BUSD maintained relative stability \cite{galati2024silicon}. These findings demonstrate that the behavior of stablecoins during crises cannot be generalized, as it varies significantly between different stablecoins and across different crisis events.

\subsection{Stablecoin and CBDC adoption}
\journalColorText{Another research question that received significant attention relates to the use cases of stablecoins, the factors influencing their adoption, and the impacts of this adoption. Stablecoins have become an essential asset on decentralized finance (DeFi) lending platforms \cite{darlin2022debt, tovanich2023contagion}. Borrowing stablecoins from these platforms often facilitates speculative trading activities \cite{rosa2021tether, wang2022speculative}. Additionally, stablecoins are increasingly used for international money transfers, providing a faster and more cost-efficient alternative to traditional financial systems. In jurisdictions where users face challenges transferring money internationally---due to sanctions, capital controls, or limited access to services---stablecoins can serve as a viable solution.  However, this raises regulatory concerns regarding their potential use for sanctions evasion \cite{travkina2022stablecoin}. Stablecoins have found many other applications, such as serving as rewards on task crowdsourcing platforms \cite{meng2023cryptocurrency}.}

\journalColorText{Despite the increasing adoption of stablecoins, several significant barriers hinder their growth. Regulatory uncertainty remains one of the most prominent challenges, discouraging adoption by both individual users and institutions \cite{adams2023investigating}. For instance, \citet{fernandez2020stable} highlighted that existing Ethereum token standards fail to meet key regulatory requirements, such as ISO 20022 compliance, which is critical for institutional payment adoption. Institutions face additional ambiguity regarding the accounting treatment of stablecoins \cite{hampl2021can}. In relation to that, \citet{hampl2021can} suggested that the low volatility of stablecoins could potentially qualify them as cash-equivalent under accounting standards; however, the ambiguity remains. Future research should work on addressing these gaps, possibly by reviewing stablecoin regulations across different jurisdictions and proposing model legal frameworks to guide policymakers. Studying the characteristics of stablecoins and proposing appropriate classifications given the current international accounting standards is also important.}


\journalColorText{Alongside these regulatory challenges, governments also express hesitations about supporting stablecoins or CBDCs due to concerns over potential disruptions to the financial system. Theoretical models suggest that widespread adoption of stablecoins or CBDCs could destabilize commercial banks by shifting deposits away from them \cite{castren2022digital} and reduce the effectiveness of central banks' open market operations \cite{park2023stablecoins}. Additionally, CBDCs could limit monetary policy autonomy by increasing interest rate synchronization across countries \cite{karau2023central}. Future research could explore methods to mitigate such risks. The case of Libra highlights the regulatory barriers to stablecoin development and adoption. Announced by Facebook in 2019, the project faced intense scrutiny from regulators, ultimately leading to its indefinite halt \cite{zetzsche2021regulating, read2020libra}.}

\journalColorText{Nevertheless, the rise of private stablecoins beyond governmental control is prompting governments worldwide to explore the introduction of their own controlled digital currencies \cite{atlanticcouncil_cbdc_tracker}. This growing interest is reflected in the significant number of studies focusing on central bank digital currencies (CBDCs). While CBDCs are not necessarily limited to being stablecoins, those examined in our study all focus on stability. This emphasis on stability aligns with the economic objectives of central banks, which aim to ensure their currencies serve as a stable store of value and a reliable medium of exchange, thereby reducing uncertainty.}

\journalColorText{Designing a CBDC, however, requires navigating complex trade-offs between central bank oversight and public trust, anonymity and legal accountability, and innovation and stability \cite{flores2023cbdc, ueda2024design, genc2024literature}. These trade-offs sometimes cause conflicting proposals. Further research exploring the optimal balance for different countries and contexts is required. For instance, researchers find that there is a clear preference for permissioned/private blockchains in CBDC design, reflecting central banks' desire for control. Private blockchains can also be more scalable and more compliant with regulations. Nevertheless, some researchers argue for decentralized designs to garner trust from investors, particularly in countries with weak institutions or a history of hyperinflation \cite{marthinsen2022hyperinflation}. Balancing user privacy with the need for AML/KYC compliance and preventing illicit activities is another significant challenge \cite{zams2020designing, flores2023cbdc, sethaput2023blockchain, genc2024literature, tronnier2021privacy}. Several research gaps exist related to this research topic, including the lack of in-depth analysis of the technical implementation and security challenges associated with CBDC designs \cite{morales2021implementing}. A significant research gap also exists in the analysis of CBDC failures. While \citet{dowd2024so} discussed several failed projects, in-depth studies investigating the reasons behind these failures are lacking.}



\subsection{Data Types}
\label{subsec:datatypes}
This section highlights and discusses the data sources researchers have relied on for stablecoin-related research. By categorizing and analyzing these data sources, we aim to provide guidance for future research.

\begin{itemize}
\item \textbf{Cryptocurrency Market Data:} This is the most common data type used by the reviewed papers. Market data mainly involves price, but it can also refer to other information, including volume and market capitalization. This type of data is easily accessible from cryptocurrency market data aggregators, such as CoinGecko \cite{coingecko} and CoinMarketCap \cite{coinmarketcap}, which provide a reliable snapshot of the cryptocurrency market at any given moment by collecting data from numerous exchanges, applying volume-weighted averages, and performing data quality checks. Cryptocurrency exchanges, such as Binance~\cite{binance_api} and Coinbase~\cite{coinbase_api}, also offer different types of market data. However, the resulting snapshot would be limited to that platform, while aggregators can collect and process data from multiple exchanges to present a broader picture of the market. 

Market data has been widely used in various areas of research. For example, many studies rely on price data to explore the relationships between stablecoins and other assets \cite{grobys2022tether, chen2022volatility, kolodziejczyk2023stablecoins}. Other studies use volume data to assess the impact of external factors, such as media announcements, on stablecoin market activity \cite{ante2021impact, saggu2022intraday}.

Researchers should also consider the granularity of the data they use. For instance, daily price data captures only one snapshot per day, which can miss intraday volatility. This is particularly important for stablecoins, where even short-term volatility can be significant. Such fluctuations can undermine market confidence or lead to liquidations in leveraged positions. Many of the studies that examined asset price relationships rely on daily price data (e.g., \cite{wang2020stablecoins, yousaf2022spillovers, ali2022examination, ghabri2022information, diaz2023stablecoins, syuhada2022tether}).

\insights{The granularity of the data is crucial. The daily data used in many studies miss intraday volatility that can undermine stablecoins.}

\item \textbf{Blockchain \journalColorText{and DeFi Platform} Data:} Blockchain data includes all the information stored on-chain, i.e., permanently added in the cryptocurrency ledger, such as transactions, smart contracts, etc.  On-chain data allows researchers to analyze the transaction network of cryptocurrencies, enabling investigations like new issuances and transfer of digital coins between users. 

For example, \citet{griffin2020bitcoin} analyzed blockchain transactions and found that stablecoin issuances impact the Bitcoin market. Many researchers also focus on stablecoin transactions during market collapses \cite{kitzler2023disentangling, briola2023anatomy, saengchote2024digital}. In particular, \citet{zhu2024data} detected manipulation patterns within stablecoin transaction data.

\journalColorText{Blockchain transaction records can also be used for governance-related analyses. Tools like Etherscan or the Maker Governance Portal provide data on governance actions, such as voter addresses, voting activity, auction data, governance token distributions, and governance token transactions \cite{sun2024decentralization}. Platforms like the MCD Voting Tracker log MakerDAO governance proposals, offering a valuable source for research related to proposals and voting. Additionally, \citet{klages2022while} analyzed liquidation data of loan collateral on MakerDAO to develop a model simulating a deleveraging spiral crisis. Researchers have also utilized data from platforms like Curve Finance, Convex Finance, and Votium to study token locks, votes, and bribe transactions \cite{lloyd2023emergent}.}

Blockchain data can be collected by deploying a full node of the cryptocurrency network, which downloads the ledger from the peer-to-peer network. However, this process is time-consuming and requires specialized knowledge of the cryptocurrency network. As a result, many researchers turn to more accessible sources, such as blockchain explorers (e.g., Etherscan \cite{etherscan}, Omni Explorer \cite{omniexplorer}, and Tronscan \cite{tronscan}), specialized data platforms (e.g., Dune Analytics \cite{dune_analytics}, The Graph Protocol \cite{the_graph}), and publicly available datasets (e.g., Ethereum ETL \cite{ethereum_etl}, Chartalist repository \cite{chartalist_repository}).

Despite their accessibility, these sources have some limitations. Researchers must place trust in blockchain explorers or other third parties to accurately extract data from the blockchain ledgers. In this regard, some researchers have pointed out inaccuracies in data from blockchain explorers \cite{caprolu2021analysis, he2023don}.

\item \textbf{Traditional Financial Data:} This type of data is used to study the relationship between traditional financial assets and stablecoin markets. Traditional financial data used in the considered papers includes market data of traditional assets (e.g., stocks \cite{wang2020stablecoins} and commodities \cite{belguith2024can}) and macroeconomic indicators (e.g., interest rates \cite{nguyen2022stablecoins} and GDP \cite{bojaj2022forecasting}).

\item \textbf{Synthetic Data:} Synthetic data has primarily been used to simulate market conditions and evaluate the performance of proposed stablecoin designs. For example, Monte Carlo simulations have been utilized by various authors for these purposes \cite{nascimento2023scaled, liu2020mover}. 

\item \journalColorText{\textbf{Surveys and Interviews:} Many researchers used data gathered from interviews or questionnaires to explore various aspects of stablecoin adoption, use cases, and perceptions. Questionnaires have been conducted with end-users to understand their preferences regarding stablecoins \cite{ante2023profiling, jin2023preferring, au2024can}. Platforms like CoinGecko \cite{ante2023profiling} and Amazon MTurk \cite{au2024can} have been used for this purpose. Beyond user-focused surveys, interviews with industry leaders and experts were conducted. For example, \citet{sood2023identification} conducted interviews with 15 blockchain and financial sector professionals, and \citet{singh2023quest} interviewed 27 policymakers, macroeconomic experts, and academics to assess the challenges and implications of stablecoin adoption. (add this sentence smoothly: \citet{morales2021implementing} and \citet{zams2020designing} conducted interviews with project members of CBDC projects and relevant experts from various countries). (add a suitable conjunction here), \citet{ayuba2022conceptual} conducted semi-structured interviews with industry experts to investigate the potential use cases of stablecoins within the construction industry. }

\item \textbf{Stablecoin Documentation:} This includes white papers, technical reports, and official documentation released by stablecoin \journalColorText{or CBDC pilot projects}. These documentations serve as a source of information for understanding the design and mechanisms of stablecoin projects. For example, several papers included a survey of such documents to categorize stablecoins based on their mechanisms \cite{mita2019stablecoin, pernice2019monetary, moin2020sok, klages2022while}. 

\item \journalColorText{\textbf{Regulatory Documents:} Regulatory documents are an important data source for studying the legal and policy landscape surrounding stablecoins and CBDCs. Researchers utilize these documents to understand the stances of governments and to conduct qualitative or comparative regulatory analyses, evaluating existing frameworks across jurisdictions and identifying the challenges regulators face in relation to stablecoins and CBDCs \cite{zetzsche2021regulating, ferreira2021curious, read2020libra, travkina2022stablecoin, shi2021legislation, schwarcz2022regulating}. National regulatory frameworks analyzed include the U.S. Stablecoin TRUST Act and the European Union’s Markets in Crypto-Assets (MiCA). Reports from institutions such as the Bank for International Settlements (BIS) \cite{adrian2021rise}, the International Monetary Fund (IMF) \cite{prasad2023will}, and various national central banks (e.g., the European Central Bank \cite{bolt2022getting}, the People's Bank of China \cite{allen2022fintech}) have also been used.}

\item \textbf{Social Media:} Some of the reviewed papers use Twitter posts to gauge market sentiment and its relationship to stablecoin markets \cite{ferretti2023using, ferretti2024cryptocurrency, ali2022examination, saggu2022intraday}. Note that this approach is sensitive to both the subset of posts analyzed and the methodology used for sentiment classification. \journalColorText{Also, using Twitter as a data source has become significantly less accessible with increased fees and reduced data access \cite{hendrix2023twitter}. Researchers can explore other social media platforms such as Reddit, Telegram, Discord, StockTwits, etc.} 

\item \textbf{News Articles:} Used in several papers \cite{ayadi2023directional, de2024dollar, osman2024economic}, news articles can be analyzed using natural language processing tools to assess their impacts on stablecoin markets. Sentiment analysis of news articles has led to the development of indices, such as the CBDC Uncertainty and CBDC Attention indices \cite{wang2022effects}, which can be directly utilized by researchers.
\end{itemize}

\section{Conclusion}
\label{sec:conclusion}

\journalColorText{The stablecoin ecosystem has rapidly become a major component of digital finance, with transaction volumes rivaling traditional payment networks. While academia has given increasing attention to stablecoins, existing research remains fragmented due to its inherently interdisciplinary nature, encompassing economic, technological, and regulatory perspectives. This work addresses this fragmentation, offering a structured and comprehensive analysis designed to synthesize existing stablecoin-related knowledge.}

\journalColorText{Through an extensive review of 239 Scopus-indexed academic papers, we categorized findings into 26 specific research questions spanning economic, technical, and policy domains. Our analysis critically evaluated methodological approaches, identified common data sources, and correlated insights to provide a unified understanding of the field. We highlighted critical areas lacking consensus or sufficient evidence, such as the disputed relationships between stablecoin issuance and cryptocurrency market dynamics, and discussed the implications of the most important findings, such as the geopolitical effects of stablecoins being predominantly pegged to the US dollar. Furthermore, we identified prevalent methodological limitations, including reliance on daily price data that overlooks intraday volatility, and identified substantial gaps in current research, particularly regarding security and privacy concerns beyond conceptual levels, vulnerabilities inherent in decentralized stablecoin governance structures, and neglected lessons from historical stablecoin failures.}

\journalColorText{By synthesizing current knowledge, critically analyzing methodological approaches, and discussing research gaps, this survey aims to guide and catalyze future research on stablecoins, an asset class whose increasing economic importance has significant implications across social, political, and technological domains.}

\bibliographystyle{IEEEtranN}
\bibliography{references}

\end{document}